\documentclass[aps,prl,twocolumn,showpacs]{revtex4-1}
\usepackage{amssymb}
\usepackage{graphicx}
\usepackage{dcolumn}
\usepackage{bm}
\usepackage{amsmath}
\usepackage{soul,color}
\usepackage[colorlinks,linkcolor=red,citecolor=blue]{hyperref}

\definecolor{cGreen}{RGB}{0,0,0}
\definecolor{cBlue}{RGB}{0,0,139}
\definecolor{cmagenta}{RGB}{0,0,0}

\begin{document}

\title{Spin-Nematic Vortex States in Cold Atoms}
\author{Li Chen$^{1,2}$}
\author{Yunbo Zhang$^{3}$}
\email{ybzhang@zstu.edu.cn}
\author{Han Pu$^{4}$}
\email{hpu@rice.edu}

\affiliation{
$^1${Institute of Theoretical Physics and State Key Laboratory of Quantum Optics and Quantum Optics Devices, Shanxi University, Taiyuan 030006, China}\\
$^2${Institute for Advanced Study, Tsinghua University, Beijing 100084, China}\\
$^3${Key Laboratory of Optical Field Manipulation of Zhejiang Province and Physics Department of Zhejiang Sci-Tech University, Hangzhou 310018, China}\\
$^4${Department of Physics and Astronomy, and Rice Center for Quantum
    Materials, Rice University, Houston, TX 77005, USA}
}

\begin{abstract}
The (pseudo-)spin degrees of freedom greatly enriches the physics of cold atoms. This is particularly so for systems with high spins (i.e., spin quantum number larger than 1/2). For example, one can construct not only the rank-1 spin vector, but also the rank-2 spin tensor in high spin systems. Here we propose a simple scheme to couple the spin tensor and the center-of-mass orbital angular momentum in a spin-1 cold atom system, and show that this leads to a new quantum phase of the matter: the spin-nematic vortex state that features vorticity in an SU(2) spin-nematic tensor subspace. Under proper conditions, such states are characterized by quantized topological numbers. Our work opens up new avenues of research in topological quantum matter with high spins.
\end{abstract}

\maketitle

\textit{Introduction ---} Cold atoms are well known to provide an ideal platform for quantum simulation \cite{Bloch2012}. As a quantum simulator, cold atoms can not only simulate important toy models arising from other subfields of physics, but also offer opportunities to construct new models that take advantages of their unique properties. One particular example is synthetic spin-orbit coupling (SOC) generated either by Raman laser coupling~\cite{Galitski2013, Goldman2014, Zhai2015, WZhang2018} or by periodic modulation \cite{Chin2019,Gorg2019,Schweizer2019}, due to the flexibility of tailoring laser configuration or Floquet engineering, novel types of SOC not naturally occur in other systems can be realized. Another unique property of the atom is that the number of internal states, (i.e., the spin) involved can be tuned to some extent, which makes possible the exploration of intriguing physics of high spins \cite{Kawaguchi2012}.

Combining SOC and high spin, SOC in cold atoms with high spins has received much attention in recent years. Raman laser induced SOC in spin-1 condensate \cite{Lan2014} was realized in the group of Spielman \cite{Campbell2016}, where various phase transitions and the associated quantum tricritical point have been identified. Very recently, interesting phenomena have been explored in a novel type of coupling between the center-of-mass orbital angular momentum (OAM) and the spin vector in spin-1 condensates \cite{Chen2016,HChen2018,PChen2018} where topological spin vortices, as well as the Hess-Fairback effect have been observed.

Nevertheless, previous works, including the studies mentioned above, predominantly focus on the textures of the spin operators but few on those of the nematic tensors \cite{Kawaguchi2012,Mueller2004,Podolsky2005}. The nematicities, which serve as fundamental quantities in high-spin quantum systems, have proved to be of wide usage in distinguishing different phases \cite{Kawaguchi2012,Podolsky2005} or generating topological structures \cite{Kawaguchi2012,Ruostekoski2003,Ueda2014}. In the current work, we propose to synthesize the coupling between the OAM and the spin-nematic tensor in a spin-1 condensate, and show that this coupling leads to a novel vortex state in a special SU(2) subspace spanned by a combination of spin and nematic operators. We call such states, which have never been studied before, the {\em spin-nematic vortex state}.

\begin{figure}[t]
	\includegraphics[width=0.46\textwidth]{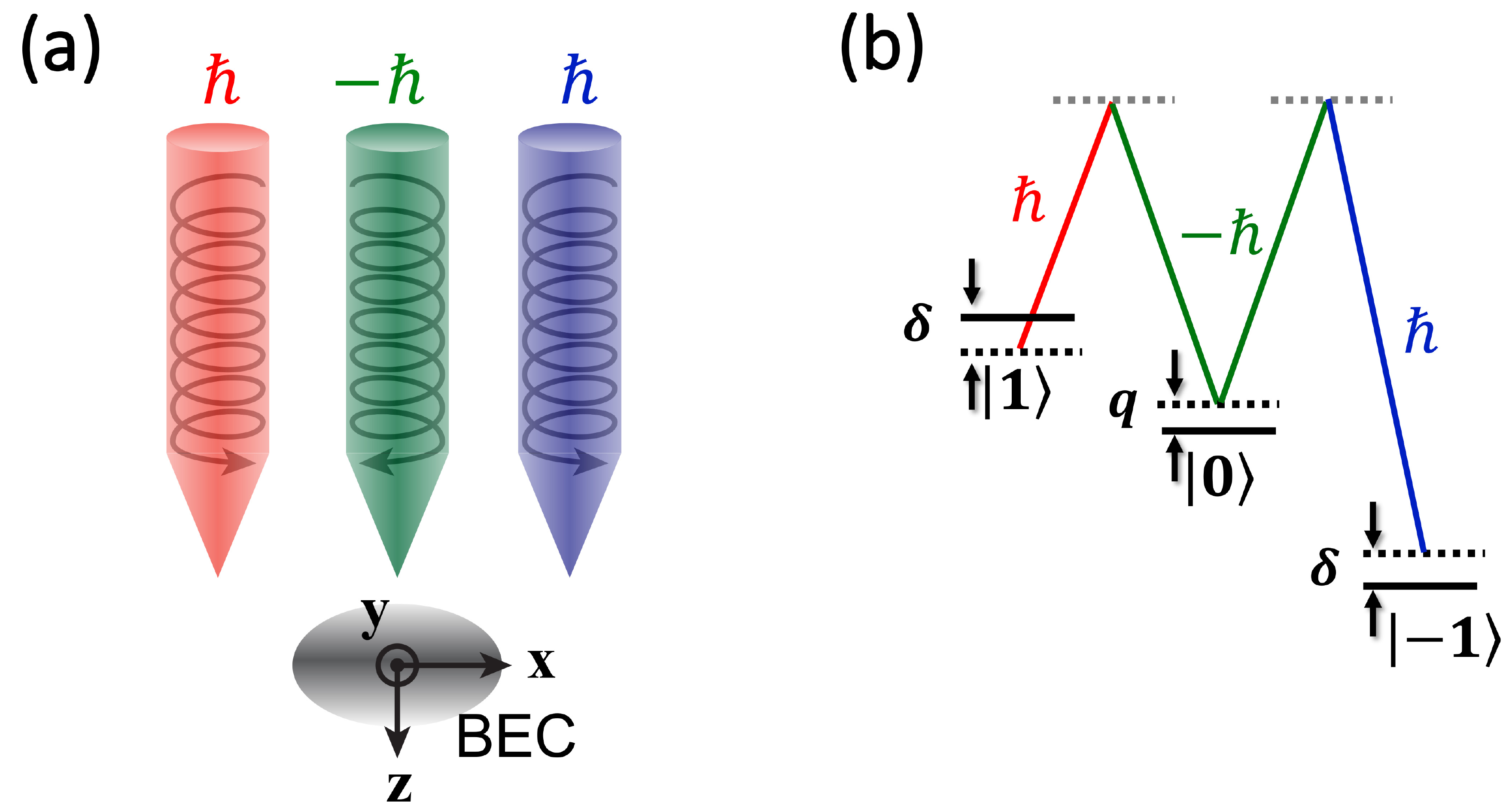}
	\caption{{\color{cmagenta}(a) Schematic of the system.} The three laser beams create a pair of Raman transitions that couple the atomic spin nematic tensor and its OAM. (b) Atomic level structure.  }
	\label{Fig1}
\end{figure}

\textit{Spin nematic and OAM coupling ---} Our proposal builds upon previous studies on coupling spin and OAM \cite{Chen2016} and on coupling spin tensor and linear momentum~\cite{Luo2017} in cold atoms. Specifically, we consider a spin-1 condensate confined in a three-dimensional harmonic trap with cylindrical symmetry, as is schematically shown in Fig.~\ref{Fig1}, where the three spin states with magnetic quantum number $m_F=1$, 0 and $-1$ are Raman coupled by three co-propagating Laguerre-Gaussian beams along the $z$-axis. Two of the laser beams carry OAM $L_z=\hbar$ and the third beam has $L_z=-\hbar$. {\color{cmagenta} The single-particle Hamiltonian can be written in the cylindrical coordinates $(r,\phi,z)$ as (taking $\hbar=m=\omega=1$ with $m$ the atomic mass, and $\omega$ the transverse trap frequency) \cite{SM}:
\begin{eqnarray}
H_{0} &=&-\frac{1}{2}\nabla ^{2}+\frac{1}{2}r^{2} +\frac{1}{2}\gamma^2z^{2} +\delta S_z+q S_{z}^2  \notag \\
&&+i\sqrt{2}\Omega _{R}\left(
r\right)\left( e^{2i\phi}\left|z\right\rangle \left\langle y \right| - e^{-2i\phi}\left|y\right\rangle \left\langle z \right|\right) ,  \label{A1}
\end{eqnarray}
where $\mathbf{S} = \{S_x,S_y,S_z\}$ are the spin operators, $\gamma = \omega_z/\omega$ the aspect ratio between the longitudinal and the transverse confinement}, {\color{cGreen}$\delta$ the two-photon Raman detuning which we take to be zero corresponding to the case of two-photon resonance}, $q$ the effective quadratic Zeeman energy, $\Omega _{R}\left( r\right) =2\Omega _{0}\left( \frac{r}{w}\right)^{2}e^{-2r^{2}/w^{2}}$ the $r$-dependent Raman coupling strength with $\Omega_0$ characterizing the Raman beam intensity, and $w$ the beam width. In Eq.~(\ref{A1}), instead of the bare spin states $|m_F=+1 \rangle$, $|0 \rangle$, and $|-1\rangle$, we have used the Cartesian polar states $|\mu \rangle$ with $\mu=x$, $y$ and $z$, which are defined as the eigenstates to spin operator $S_\mu$ with zero eigenvalue, i.e., $S_\mu |\mu \rangle =0$. In terms of the bare spin states (i.e., eigenstates of $S_z$), we have $\left|x\right\rangle = \frac{1}{\sqrt{2}}\left( \left|-1\right\rangle - \left|1\right\rangle\right),
\left|y\right\rangle = \frac{i}{\sqrt{2}}\left( \left|-1\right\rangle + \left|1\right\rangle\right)$, and $
\left|z\right\rangle = \left|0\right\rangle$ \cite{Ohmi1998}. {\color{cmagenta}Furthermore, since the Raman induced coupling (last term in Eq.~(\ref{A1})) are only in the transverse $r$-$\phi$ plane, we can take our computation in the transverse plane \cite{SM}. This effective reduction of spatial dimension allows us to significantly increase our computation efficiency without losing any essential physics.}

The coupling between nematic tensor and OAM in $H_0$ can be more clearly seen when we carry out a gauge rotation $U=\exp(2i\phi S_z^2)$. The new single-particle Hamiltonian $\tilde{H}_0 = U H_0 U^\dagger$ in the rotating frame is given by
\begin{equation}
\begin{aligned}
\tilde{H}_{0} &=\frac{\left(i\nabla - \mathbf{A}\right)^2}{2}+\frac{r^{2}}{2}+q S_{z}^2+\sqrt{2}\Omega _{R}S_x\\
&= -\frac{\nabla^2}{2}-\frac{2\left(L_z-S_z^2\right)S_z^2}{r^2}+\frac{r^{2}}{2}+q S_{z}^2+\sqrt{2}\Omega _{R}S_x,
\end{aligned}
\label{A2}
\end{equation}
where $\mathbf{A} = -iU^\dagger\nabla U=2S_z^2\hat{e}_\phi/r$ is the synthetic gauge field on the azimuthal direction $\hat{e}_\phi$. In Eq.~(\ref{A2}), one can clearly see the spin-nematic-OAM coupling term $\sim L_z S_z^2$ which couples the atomic quasi-OAM $L_z = -i\partial_\phi$ with one of the irreducible nematic tensors $N_{zz} = S_z^2-2/3$. It will play a crucial role in inducing various spin-nematic vortex states.

\textit{Single-particle properties ---} We investigate the spectrum and the eigenstates of $\tilde{H}_0$. First, we realize that $L_z$ is conserved as $[L_z,\tilde{H}_0]=0$. Furthermore, both $L_z$ and $\tilde{H}_0$ commute with the spin parity operator
\begin{equation}
P=|+1 \rangle \langle -1 | + |-1 \rangle \langle +1 | +|0 \rangle \langle 0 |  \,,\label{A3}
\end{equation}
which satisfies $P^2=1$. $P$ carries a pair of eigenvalues $P=\pm1$ distinguishing spin parity of the eigenstates. In particular, the states with even parity ($P=+$) possess the same phase on the ${\pm1}$ spin components, whereas those with odd parity ($P=-$) have a phase difference of $\pi$ on the ${\pm1}$ components. It is straightforward to see that the Cartesian states $|x\rangle$ has odd spin parity, while $|y \rangle$ and $|z \rangle$ possess even spin parity. The conservation of $L_z$ ensures the quasi-OAM $l_z$ to be a good quantum number in the rotating frame, and hence the energy eigenstates can be labeled using $P$ and $l_z$ as  $\tilde{\mathbf{\Psi}}_{P=\pm,l_z}=e^{i l_z \phi}\boldsymbol{\xi}_{\pm}(r)=e^{i l_z \phi}\left[\xi_1(r),\xi_0(r),\xi_{-1}(r)\right]^T$. The spinor wave function $\boldsymbol{\xi}_\pm(r)$ can be expanded in the Cartesian basis as
\begin{equation}
\begin{aligned}
\boldsymbol{\xi}_{+} &= \sqrt{\rho(r)} \,\left[i\cos{\Theta(r)}\left|y\right\rangle +\sin{\Theta(r)}\left|z\right\rangle\right],\\
\boldsymbol{\xi}_{-} &= \sqrt{\rho(r)}\,\left|x\right\rangle,
\end{aligned}
\label{A4}
\end{equation}
where $\rho(r) = |\tilde{\boldsymbol{\Psi}}(r)|^2$ is the total particle density, and $\Theta(r)$ characterizes the $r$-dependent superposition weight.

\begin{figure}[t]
\includegraphics[width=0.48\textwidth]{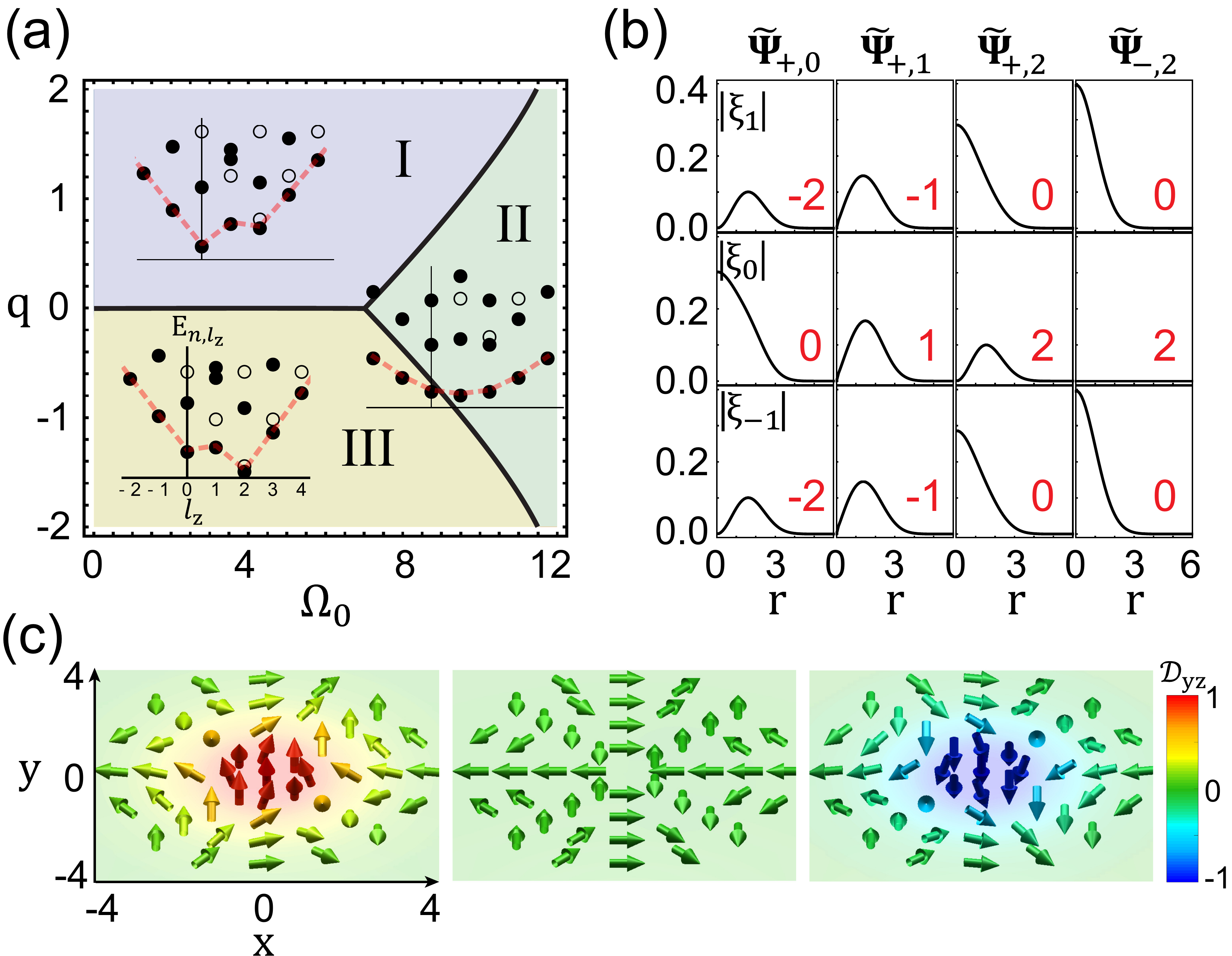}
\caption{(a) Single-particle phase diagram with solid lines indicating first-order phase transitions. Insets: typical dispersion spectra at $(\Omega_0=4,q=0.5)$, $(10,0)$, and $(4,-0.5)$ in three parametric regimes I, II, and III. Solid dots and hollow circles denote the states with even and odd parity, respectively. {\color{cGreen}(b) Spinor wave functions of typical lower-lying eigenstates $\tilde{\mathbf{\Psi}}_{P=+,l_z=0,1,2}$ and $\tilde{\mathbf{\Psi}}_{P=-,l_z=2}$.} Red numbers display the mechanical OAM on each spin component in the lab frame. (c) Typical spin-nematic textures for Phase I, II and III (from left to right). Arrows represent the spin-nematic vector $\boldsymbol{\mathcal{Q}}=\{\mathcal{S}_x,2\mathcal{N}_{yz},\mathcal{D}_{yz}\}$, and the arrow color indicates the strength of $\mathcal{D}_{yz}$.}
\label{Fig2}
\end{figure}

We numerically solve the Schr{\"o}dinger equation to obtain the energy spectrum and the eigenstates. Figure~\ref{Fig2}(a) displays the single-particle ground-state phase diagram in the parameter space spanned by $\Omega_0$ and $q$. Three phases I, II and III can be identified. The ground states in all three phases have even spin parity, while their quasi-OAM quantum numbers $l_z$ are 0, 1, and 2, respectively. All the phase transitions in the diagram Fig.~\ref{Fig2}(a) are of first-order since, across the phase boundary, the first order derivative of the ground-state energy with respect to $\Omega_0$ or $q$ exhibit discontinuity \cite{SM}. In each phase, we show typical energy spectrum as the inset in Fig.~\ref{Fig2}(a), where the solid dots and hollow circles distinguish the even and odd spin parity states. In Fig.~\ref{Fig2}(b), we plot the magnitude of the wave function for each bare spin component $|\xi_{0,\pm1}|$. The first three columns represent the ground state in each phase, and the last column corresponds to an odd spin parity state $\tilde{\mathbf{\Psi}}_{-,2}$ in Phase III that lies very close to the ground state. This state will be important in our later discussion on the many-body effects in a weakly interacting condensate.

The case with vanishing quadratic Zeeman term (i.e., $q=0$) deserves some special attention. Under this situation, the single-particle spectrum of the even-parity states is symmetric about $l_z=1$ \cite{SM}. For $\Omega_0$ smaller than a critical value $\Omega_c\approx 7$, the line $q=0$ represents the boundary between Phase I and III [see Fig.~\ref{Fig2}(a)], on which the states $\tilde{\mathbf{\Psi}}_{+,l_z}$ with $l_z=0$ and 2 are degenerate. When $\Omega_0 > \Omega_c$, we enter into Phase II with $l_z=1$. Since the term $\left(L_z-S_z^2\right)S_z^2$ in Hamiltonian~(\ref{A2}) vanishes for $l_z=1$ due to $S_z^2=S_z^4$, $S_x$ is now a conserved quantity. As a result, the spinor wave function $\boldsymbol{\xi}_{+}$ becomes an eigenstate of $S_x$ featuring $\Theta=\pi/4$.

\textit{Spin-nematic vortices --- }
In the lab frame, different spin components of the single-particle states carry different mechanical OAM as is indicated by the numbers in the subplots of Fig.~\ref{Fig2}(b). For the spin-0 component, this is simply the quasi-OAM quantum number $l_z$; whereas for the spin-$(\pm 1)$ component, it is $l_z -2$. The transformation between the wave function in the lab frame $\mathbf{\Psi}$ and that in the rotating frame $\tilde{\mathbf{\Psi}}$ is given by $\mathbf{\Psi} = U^\dagger\mathbf{\tilde{\Psi}}$, which explicitly leads to
\begin{equation}
\begin{aligned}
\mathbf{\Psi}_{+,l_z} &= \sqrt{\rho(r)}e^{i l_z \phi}\left(ie^{-2i \phi}\cos{\Theta(r)}\left|y\right\rangle +\sin{\Theta(r)}\left|z\right\rangle\right), \\
	\mathbf{\Psi}_{-,l_z} &= \sqrt{\rho(r)}e^{i (l_z-2) \phi}\left|x\right\rangle.  \end{aligned}
\label{A5}
\end{equation}
Clearly, the ground state of Phase II, represented by $\mathbf{\Psi}_{+,l_z=1}$, is a singular vortex as each of its bare spin components carry a finite vorticity and hence the total density vanishes at $r=0$; while those in Phase I and III, represented by $\mathbf{\Psi}_{+,l_z=0,2}$, are coreless vortices that contain at least one spin component with no vorticity with finite density at $r=0$.

We investigate the spin and nematic textures by calculating the normalized spin density
\begin{equation}
	\mathcal{S}_\mu=\frac{\mathbf{\Psi}^\dagger S_\mu \mathbf{\Psi}}{\rho(r)},
	\label{A6}
\end{equation}
and the normalized nematic density
\begin{equation}
	\mathcal{N_{\mu\nu}}=\frac{\mathbf{\Psi}^\dagger N_{\mu\nu} \mathbf{\Psi}}{\rho(r)},
	\label{A7}
\end{equation}
where $N_{\mu\nu} = \frac{1}{2}\left(S_\mu S_\nu+S_\nu S_\mu\right)-\frac{2}{3}\delta_{\mu\nu}$ are the symmetrized SU(3) nematic tensors with nine components by taking $\mu,\nu=x,y,z$ \cite{Kawaguchi2012, Mueller2004, Podolsky2005}. Diagonalizing the nematic density matrix $\mathcal{N}$ results in three eigenvalues $\lambda_{1,2,3}$ characterizing the alignment axis of nematic orders. A uniaxial nematic state is characterized by $\lambda_1\neq\lambda_2=\lambda_3$, while for a biaxial nematic state, none of these eigenvalues are equal. We remark that the Cartesian states are closely related to the SU(3) operators, which greatly facilitates the calculation of $\boldsymbol{\mathcal{S}}$ and $\mathcal{N}$ \cite{SM}.

Now we discuss the four low-energy states $\mathbf{\Psi}_{+,l_z=0,1,2}$ and $\mathbf{\Psi}_{-,l_z=2}$ represented in Fig.~\ref{Fig2}(b). The odd-parity state $\mathbf{\Psi}_{-,2}$ is topologically trivial, since it is simply a polar state with vanishing spin density, and a fixed uniaxial direction along $N_{xx}$. For the even-parity states $\mathbf{\Psi}_{+,l_z=0,1,2}$ which represent the ground state in the three phases, we have
\begin{equation}
\begin{aligned}
\boldsymbol{\mathcal{S}}&=\{-\cos(2\phi)\sin(2\Theta),0,0\}, \\
\mathcal{N}&=
\begin{bmatrix}
\frac{1}{3} & 0 & 0 \\
0 & -\frac{1}{6}\left(1+3\cos 2\Theta\right) & -\frac{1}{2}\sin(2\Theta)\sin(2\phi) \\
0 & -\frac{1}{2}\sin(2\Theta)\sin(2\phi) & -\frac{1}{6}\left(1-3\cos 2\Theta\right)
\end{bmatrix},
\end{aligned}
\label{A8}
\end{equation}
where the spin vector $\boldsymbol{\mathcal{S}}$ is polarized along $S_x$. The eigenvalues of the nematic matrix $\mathcal{N}$ can be obtained as $\lambda_1 = 1/3$, $\lambda_{2,3} = -1/6[1\pm3\sqrt{1-\sin^2(2\Theta)\cos^2(2\phi)}]$, indicating that all three states are biaxial nematic states. Furthermore, the spin and nematic densities in Eq.~(\ref{A8}) are not independent, and satisfy \cite{Mueller2004}
\begin{eqnarray}
\frac{1}{2}\left|\boldsymbol{\mathcal{S}}^2\right| + \text{Tr}[\mathcal{N}^2] &= & \frac{2}{3}.
\label{A9}
\end{eqnarray}
Substituting Eq.~(\ref{A8}) into Eq.~(\ref{A9}), one immediately obtains a relation
\begin{equation}
	\mathcal{S}_x^2+\mathcal{D}_{yz}^2 + (2\mathcal{N}_{yz})^2 = 1,
	\label{A10}
\end{equation}
where $\mathcal{D}_{yz} = \mathcal{N}_{yy} - \mathcal{N}_{zz} = -\cos(2\Theta)$. This relation motivates the construction of the following spin-nematic vector
\begin{equation}
\boldsymbol{\mathcal{Q}}=\{\mathcal{S}_x,\, 2\mathcal{N}_{yz},\, \mathcal{D}_{yz}\},
\label{A10_2}
\end{equation}
 which forms vector space lying on a unit Bloch sphere. In fact, the vector $\boldsymbol{\mathcal{Q}}$ is defined on an SU(2) group generated by $\mathbf{Q}=\{S_x,2N_{yz},D_{yz} = N_{yy}-N_{zz}\}$. Mathematically, the group $\mathbf{Q}$ is a type-2 subgroup of the SU(3) Lie group with the structure constant being equal to 2 \cite{SM,Yukawa2013}, i.e. $[2N_{yz},S_x]=2iD_{yz}$.

In Figure~\ref{Fig2}(c), we display the textures of the spin-nematic vector $\boldsymbol{\mathcal{Q}}$ for the ground states of the three phases $\mathbf{\Psi}_{+,l_z=0,1,2}$. One can clearly see that the transverse components of $\boldsymbol{\mathcal{Q}}$ for all three states form a vortex pattern, hence the name {\em spin-nematic vortex states}.

{\color{cGreen}We now investigate the topological properties of the spin-nematic vortex states. Let us first focus on the $q=0$ case. The ground state in Phase II is a singular vortex state $\mathbf{\Psi}_{+,l_z=1}$ satisfying $\Theta=\pi/4$ as mentioned before. Hence $\boldsymbol{\mathcal{Q}}$ is a planar vector lies on the equator of the Bloch sphere since its $z$-component vanishes, i.e., $\mathcal{D}_{yz} = -\cos(2\Theta) =0$. This allows us to depict the singular vortex $\mathbf{\Psi}_{+,l_z=1}$ by the winding number, which counts the winding times of $\boldsymbol{\mathcal{Q}}$ as one walks along a closed loop in the $x$-$y$ plane. Apparently, the winding number of the singular vortex $\mathbf{\Psi}_{+,l_z=1}$ equals to $-2$ if the loop encloses the vortex core (at the origin $x=y=0$), and equals to $0$ otherwise.
For the two coreless states $\mathbf{\Psi}_{+,l_z=0,2}$ of Phase I and III, the spin-nematic vector $\boldsymbol{\mathcal{Q}}$ is defined on the whole Bloch sphere. However, we have the boundary condition $\Theta(r\rightarrow\infty)=\pi/4$, which can be clearly observed as the contribution of term $\left(L_z-S_z^2\right)S_z^2/r^2$ in Hamiltonian~(\ref{A2}) diminishes as $r\rightarrow\infty$. Consequently, each state ($\mathbf{\Psi}_{+,l_z=0}$ or $\mathbf{\Psi}_{+,l_z=2}$) covers one half of the Bloch sphere, and their topological number can thus be characterized by the skyrmion number \cite{Nagaosa2013} defined as
\begin{equation}
\mathcal{W} = \frac{1}{4\pi} \iint d^2{r} \, \boldsymbol{\mathcal{Q}}\cdot (\partial_x \boldsymbol{\mathcal{Q}}\times \partial_y \boldsymbol{\mathcal{Q}})\,,
\label{A11}
\end{equation}
which turns out to be $\mp1$ for states $\mathbf{\Psi}_{+,l_z=0,2}$, respectively \cite{footnote}. Note that the coreless spin-nematic vortices defined in the type-2 subspace $\mathbf{Q}$ here are analogous to the Mermin-Ho vortex \cite{Mermin1976, Mermin1979, Makela2003} defined in the type-1 subspace $\mathbf{S}=\{S_x,S_y,S_z\}$. However, the two are not mathematically equivalent since the subspaces $\mathbf{S}$ and $\mathbf{Q}$ cannot be transformed into each other by SU(3) rotations \cite{Yukawa2013}. We also note that, these topological numbers are not well defined at finite $q$. For example, at finite $q$, the singular vortex can no longer be described by the winding number since $\mathbf{Q}$ is no longer restricted in the $S_x$-$2{N}_{yz}$ plane, and the skyrmion number of the other two coreless vortices are also in general not quantized to be integers.}

\begin{figure}[t]
\includegraphics[width=0.46\textwidth]{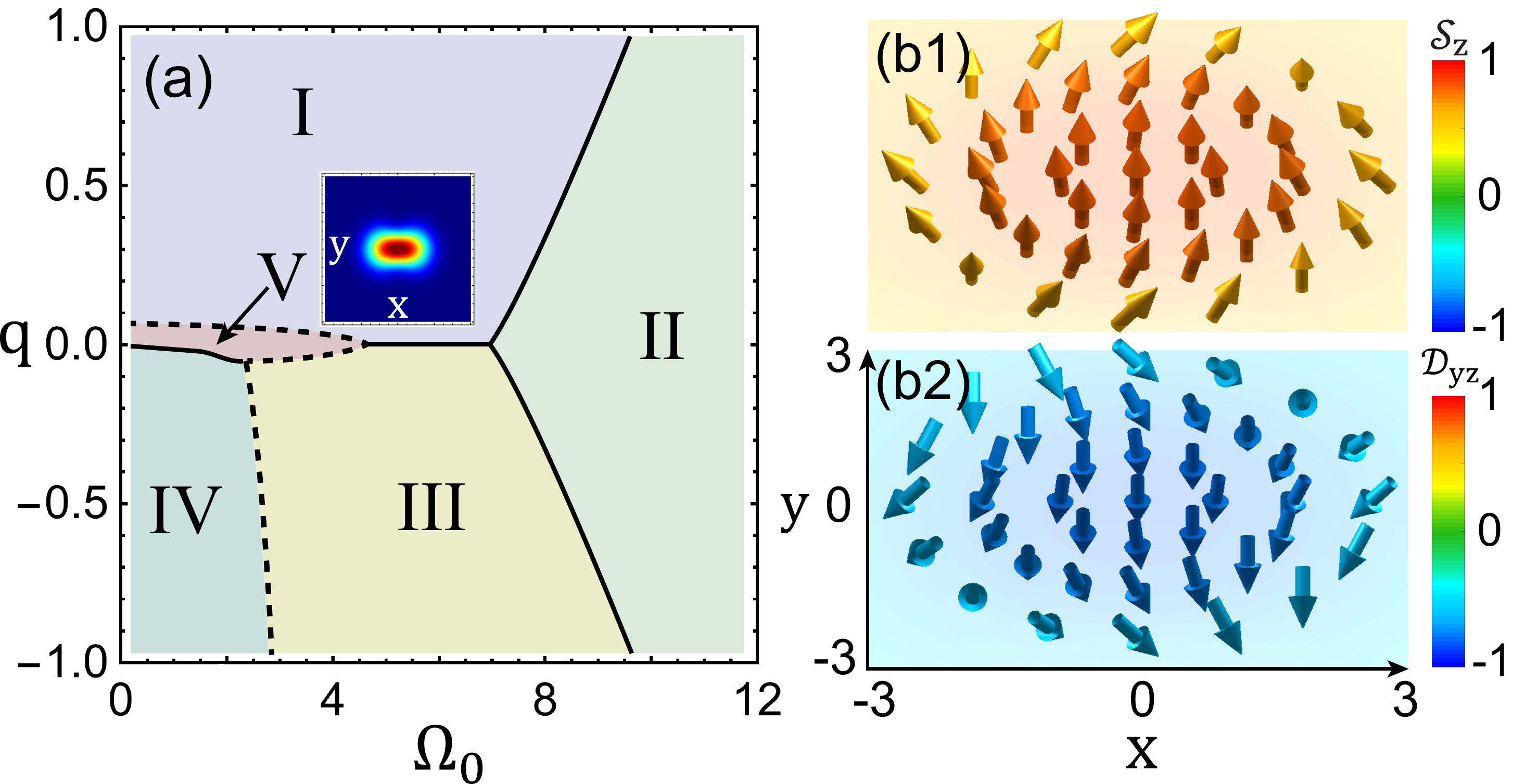}
\caption{(a) Many-body phase diagram, where the solid and the dashed lines indicate the first- and the second-order phase transitions, respectively. Insets: total density profile $\rho(r)$ of phase V. (b1) and (b2) display the typical spin texture $\boldsymbol{\mathcal{S}}=\{\mathcal{S}_x,\mathcal{S}_y,\mathcal{S}_z\}$ and spin-nematic texture $\boldsymbol{\mathcal{Q}}=\{\mathcal{S}_x,2\mathcal{N}_{yz},\mathcal{D}_{yz}\}$ of the phase IV, respectively. {\color{cGreen} The arrow color in subfigures (b1) and (b2) indicate the $z$ components of the spin density $\boldsymbol{\mathcal{S}}$ and spin-nematic density $\boldsymbol{\mathcal{Q}}$, respectively.}}
\label{Fig3}
\end{figure}

\textit{Many-body effects ---}
Next, we consider a weakly interacting condensate and discuss the effects of the interactions under the framework of the mean-field theory. In the lab frame, the interacting Hamiltonian of the spin-1 condensate is in the well-known form of \cite{Kawaguchi2012,Stamper-Kurn2013}
\begin{equation}
H_{\rm int}=\frac{1}{2} \int d^{2}r \, \rho^{2}\left( {\bf r}\right) \left[ c_{0} +c_{2}\boldsymbol{\mathcal{S}}%
^{2}\left( {\bf r}\right) \right] ,
\label{A12}
\end{equation}
where $c_0$ and $c_2$ denote the strength of the density-density and the spin-exchange interactions, respectively. We implement two different methods to obtain the mean-field ground states --- the variational method and the numerical method by solving 3D Gross-Pitaevskii equations \cite{SM}, and the results turn out to be in excellent agreement. For the variational method, we assume the condensate wave function is a linear combination of the four lower-lying single-particle states shown in Fig.~\ref{Fig2}(b):
\begin{equation}
\tilde{\mathbf{\Psi}} = D_0 \tilde{\mathbf{\Psi}}_{+,0} + D_1 \tilde{\mathbf{\Psi}}_{+,1} + D_{+} \tilde{\mathbf{\Psi}}_{+,2} +  D_{-} \tilde{\mathbf{\Psi}}_{-,2},
\label{A13}
\end{equation}
where $D_{j=0,1,+,-} = |D_j|e^{i\theta_j}$ are variational amplitudes satisfying $\sum_j |D_j|^2= 1$, {\color{cGreen} with $\theta_j$ being the phase angle of $D_j$.} We obtain the ground states by minimizing the total energy functional with respect to $D_j$.
Here, we consider a weak ferromagnetic spin-exchange interaction by taking $c_0=1$ and $c_2=-0.1 c_0$, and then map out the ground-state phase diagram as is displayed in Fig.~\ref{Fig3}(a).

The main structure of the many-body phase diagram Fig.~\ref{Fig3}(a) are consistent with that of the single-particle phase diagram Fig.~\ref{Fig2}(a). In Fig.~\ref{Fig3}(a), the three phases I, II, and III are very similar to the corresponding single-particle ones in diagram Fig.~\ref{Fig2}(a) with $|D_{j=0,1,+}|=1$ in ansatz (\ref{A13}), respectively, and the phase transitions among them are all of first order (solid lines). There are, however, two new many-body phases labeled as IV and V that have no counterparts in the single-particle phase diagram, and the phase transitions related to them can be either first- (solid lines) or second-order (dashed lines), as determined by whether the first- or second-order derivatives of the ground state energy with respect to the parameters ($\Omega$ or $q$) exhibit discontinuity or not \cite{SM}.

These two new phases, IV and V, spontaneously break the spin parity and the rotation symmetry, respectively.
Specifically, the wave function of phase IV is a superposition of states $\tilde{\mathbf{\Psi}}_{\pm,2}$ with variational amplitudes satisfying $|D_\pm| \neq 0$ and $\theta_+ - \theta_- = 0$ or $\pi$ (mod[$2\pi$]). Thus, it keeps the rotational symmetry but breaks the spin-parity symmetry. Interestingly, this state exhibits vorticity in both the spin and the spin-nematic subspaces $\mathbf{S}$ and $\mathbf{Q}$ simultaneously, as are shown in Fig.~\ref{Fig3}(b1) and (b2) respectively. The breaking of the spin parity symmetry is manifested in the fact that $S_z$ is finite, i.e., unequal occupation on the bare spin-($\pm 1$) components.

The wave function of phase V is a superposition of states $\tilde{\mathbf{\Psi}}_{+,0}$ and $\tilde{\mathbf{\Psi}}_{+,2}$ with $|D_{0,+}| \neq 0$. Thus, this state maintains the spin parity symmetry, but breaks the rotational symmetry, which leads to an interesting angular striped phase \cite {Li2017,Leonard2017}. We show the total density profile $\rho(r)$ of phase V as an inset in Fig.~\ref{Fig3}(a), where the lack of the rotational symmetry is obvious.

\textit{Experimental observation ---}
Finally, let us briefly discuss the experimental detection of the spin-nematic vortex states, which can be performed either directly or indirectly. The indirect observation is to detect such features of the wave functions $\boldsymbol{\Psi}_{+,l_z = 0,1,2}$ (presented in Fig.~\ref{Fig2}~(b)) as the core structures or the mechanical-OAM numbers. Specifically, the core structure can be obtained by the spin-selected absorption imaging; the mechanical-OAM quantum numbers can be deduced from the interference pattern after different spin components are mixed with each other by a radio-frequency $\pi/2$ pulse \cite{PChen2018,HChen2018}. In contrast, the direct observation is to detect the spin-nematic textures $\boldsymbol{\mathcal{Q}}$ directly. Since the direct observation of the spin texture $\boldsymbol{\mathcal{S}}$ has been realized via the spin-sensitive dispersive imaging in a few years ago \cite{Sadler2006,Higbie2005}, this technique can be easily generalized to measure $\boldsymbol{\mathcal{Q}}$ as the nematic operators $2 N_{yz}$ and $D_{yz}$ are rotated into the measurable direction $S_x$. Practically, this rotation can be achieved by pulsing a quadratic Zeeman magnetic field which lets $\mathbf{Q}$ evolve under the government of $\sim S_z^2$ \cite{Chang2005}, or more feasibly introducing an additional far off-resonant microwave on certain Zeeman level \cite{Hamley2012}.

\textit{Summary ---}
We have proposed a scheme to couple the atomic OAM and the nematic tensor in a spin-1 cold atomic system. The ground state exhibits vorticity in a special spin-nematic subspace. Under zero quadratic field, the spin-nematic vortices can be characterized by quantized topological numbers. These features survive in the presence of weak interaction. However, the interaction may induce spontaneous symmetry breakings, and leads to a rich many-body phase diagram. Considering the spin-OAM coupling has been realized by two experimental groups very recently \cite{HChen2018,PChen2018,Zhang2018}, we expect this work to stimulate more investigations on the spin-OAM coupled quantum gases with higher spins and nematic orders.

\begin{acknowledgments}
LC would like to thank Hui Zhai for inspiring discussion on the topological features of the vortices. LC acknowledge supports from the NSF of China (Grant No. 11804205), and the Beijing Outstanding Young Scientist Program hold by H. Zhai; YZ acknowledge supports from the NSF of China (Grant No. 11674201); HP acknowledges supports from the US NSF and the Welch Foundation (Grant No. C-1669).

\end{acknowledgments}


\begin{thebibliography}{99}
\bibitem{Bloch2012}  I. Bloch, J. Dalibard, S. Nascimb\`{e}ne, Nat. Phys. \textbf{8}, 267 (2012).

\bibitem{Galitski2013} V. Galitski and I. B. Spielman, Nature \textbf{494}, 49 (2013).

\bibitem{Goldman2014} N. Goldman, G. Juzeli\={u}nas, P. \"{O}hberg, and I.
  B. Spielman, Rep. Prog. Phys. \textbf{77}, 126401 (2014).

\bibitem{Zhai2015} H. Zhai, Rep. Prog. Phys. \textbf{78}, 026001 (2015).

\bibitem{WZhang2018} Synthetic Spin-Orbit Coupling in Cold Atoms, edited by W. Zhang, W. Yi, and C. A. R. Sá Melo (World Scientific, Singapore, 2018).

\bibitem{Chin2019} C. Chin, Nat. Phys. \textbf{15}, 1106 (2019).

\bibitem{Gorg2019} F. G\"{o}rg, K. Sandholzer, J. Minguzzi, R. Desbuquois, M. Messer, T. Esslinge, and M. Aidelsburger, Nat. Phys. \textbf{15}, 1161 (2019).

\bibitem{Schweizer2019} C. Schweizer, F. Grusdt, M. Berngruber, L. Barbiero, E. Demler, N. Goldman, I. Bloch, M. Aidelsburger, Nat. Phys. \textbf{15}, 1168 (2019).

\bibitem{Kawaguchi2012} K. Kawaguchi, and M. Ueda, Phys. Rep. \textbf{520}, 253 (2012).

\bibitem{Lan2014} Z. H. Lan, and P. \"{O}hberg, Phys. Rev. A. \textbf{89}, 023630 (2014); S. S. Natu, X. P. Li, and W. S. Cole, ibid. \textbf{91}, 023608 (2015); K. Sun, C. Qu, Y. Xu, Y. Zhang, and C. Zhang, ibid. \textbf{93}, 023615 (2016); Z.-Q. Yu, ibid. \textbf{93}, 033648 (2016); G. I. Martone, F. V. Pepe, P. Facchi, S. Pascazio, and S. Stringari, Phys. Rev. Lett. \textbf{117}, 125301 (2016).

\bibitem{Campbell2016} D. L. Campbell, R. M. Price, A. Putra, A. Vald\'{e}s-Curiel, D. Trypogeorgos, and I. B. Spielman, Nat. Commun. \textbf{7}, 10897 (2016).

\bibitem{Chen2016} L. Chen, H. Pu, and Y. Zhang, Phys. Rev. A \textbf{93}, 013629 (2016).

\bibitem{HChen2018} H.-R. Chen, K.-Y. Lin, P.-K. Chen, N.-C. Chiu,   J.-B. Wang, C.-A. Chen, P.-P. Huang, S.-K. Yip, Y.  Kawaguchi, and Y.-J. Lin,Phys. Rev. Lett. \textbf{121}, 113204 (2018).

\bibitem{PChen2018} P.-K. Chen, L.-R. Liu, M.-J. Tsai, N.-C. Chiu, Y. Kawaguchi, S.-K. Yip, M.-S. Chang, and Y.-J. Lin, Phys. Rev. Lett. \textbf{121}, 250401 (2018).

\bibitem{Mueller2004} E. J. Mueller, Phys. Rev. A \textbf{69}, 033606 (2004).

\bibitem{Podolsky2005} D. Podolsky and E. Demler, New J. Phys. \textbf{7}, 59 (2005)

\bibitem{Ruostekoski2003} J. Ruostekoski and J. R. Anglin, Phys. Rev. Lett. \textbf{91}, 190402 (2003).

\bibitem{Ueda2014} M. Ueda, Rep. Prog. Phys. \textbf{77}, 122401 (2014).

\bibitem{Luo2017} X.-W. Luo, K. Sun, and C. Zhang, Phys. Rev. Lett. \textbf{119}, 193001 (2017).

\bibitem{SM} See Supplemental Materials for more information on the derivation of $H_0$, the symmetry of the single-particle spectrum at $q=0$, the SU(3) algebras and calculation details of the spin and nematic orders, phase diagrams and quantum phase transitions, and the calculation details of the 3D Gross-Pitaevskii equations, which includes Refs. \cite{Schmidt2016,Lin2011,Stenger1998,Hamley2008,Bao2013}.

\bibitem{Schmidt2016} F. Schmidt, D. Mayer, M. Hohmann, T. Lausch, F. Kindermann, and A. Widera, Phys. Rev. A \textbf{93}, 022507 (2016).

\bibitem{Lin2011} Y.-J. Lin, K. Jim\'{e}nez-Garc\'{\i}a, and I. B. Spielman, Nature (London). \textbf{471}, 83 (2011).

\bibitem{Stenger1998} J. Stenger, S. Inouye, D.M. Stamper-Kurn, H.-J. Miesner, A.P. Chikkatur, W. Ketterle, Nature \textbf{396}, 345 (1998). 

\bibitem{Hamley2008} C. D. Hamley, E. M. Bookjans, G. Behin-Aein, P. Ahmadi, M. S. Chapman, Phys. Rev. A \textbf{79}, 23401 (2008).

\bibitem{Bao2013} W. Bao, and Y. Cai, Kinet. Relat. Mod. \textbf{6}, 1-135 (2013).

\bibitem{Ohmi1998} T. Ohmi, K. Machida, J. Phys. Soc. Japan \textbf{67}, 1822 (1998).

\bibitem{Yukawa2013} E. Yukawa, M. Ueda, and K. Nemoto, Phys. Rev. A \textbf{88}, 033629 (2013).

\bibitem{Nagaosa2013} N. Nagaosa and Y. Tokura, Nat. Nano. \textbf{8}, 899 (2013).

\bibitem{footnote} {\color{cGreen} At $q = 0$, the two coreless vortex states $\mathbf{\Psi}_{+,l_z=0,2}$ are energetically degenerate in the single-particle level. Hence any linear superposition of these two states remain as an eigenstate of the single-particle Hamiltonian. These two states have well-defined winding numbers, but not their superpositions. However, this issue can be avoided as a weak density-density interaction is included into the system. Since these two states feature identical total density profile, the presence of the density-density interaction ($c_0$ term) does not break their symmetry. However, any superposition of these two states would lead to spatial stripes with higher peak densities, and hence an excessive interaction energy. Thus the interaction term favors either $\mathbf{\Psi}_{+,l_z=0}$ or $\mathbf{\Psi}_{+,l_z=2}$, but not their superposition. For details, please see Supplementary Materials \cite{SM}. }

\bibitem{Mermin1976} N. D. Mermin and Tin-Lun Ho, Phys. Rev. Lett. \textbf{36}, 594(1976).

\bibitem{Mermin1979} N. D. Mermin, Rev. Mod. Phys. \textbf{51}, 591, (1979).

\bibitem{Makela2003} H. Makela, Y. Zhang, and K.-A. Suominen, J. Phys. A: Math. Gen. \textbf{36}, 8555 (2003).

\bibitem{Stamper-Kurn2013} D. M. Stamper-Kurn and M. Ueda, Rev. Mod. Phys. \textbf{85}, 1191 (2013).

\bibitem{Li2017} J.-R. Li, J. Lee, W. Huang, S. Burchesky, B. Shteynas, F. Cagri Top, A. O. Jamison, and W. Ketterle, Nature \textbf{543}, 91 (2017).

\bibitem{Leonard2017} J. Léonard, A. Morales, P. Zupancic, T. Esslinger, and Tobias Donner, Nature \textbf{543}, 87 (2017).

\bibitem{Higbie2005}  J. Higbie, L. Sadler, S. Inouye, A. P. Chikkatur, S. R. Leslie, K. L. Moore, V. Savalli,and  D.  M.  Stamper-Kurn,  Phys. Rev. Lett. \textbf{95}, 050401, (2005).

\bibitem{Sadler2006} L. Sadler, J. Higbie, S. R. Leslie, M. Vengalattore, and D.  M.  Stamper-Kurn, Nature \textbf{443}, 312 (2006).

\bibitem{Chang2005} M.-S. Chang, Q. Qin, W. Zhang, and M. S. Chapman, Nature Physics \textbf{1}, 111 (2005).

\bibitem{Hamley2012} C. D. Hamley, C. S. Gerving, T. M. Hoang, E. M. Bookjans, and M. S. Chapman, Nature Phys. \textbf{8}, 305 (2012).

\bibitem{Zhang2018} D. Zhang, T. Gao, P. Zou, L. Kong, R. Li, X. Shen, X.-L. Chen, S.-G. Peng, M. Zhan, H. Pu, and K. Jiang, Phys. Rev. Lett. \textbf{122}, 110402 (2019).

\end{thebibliography}
\end{document}


\title{Supplemental Materials: Spin-Nematic Vortex States in Cold Atoms}
\author{Li Chen$^{1,2}$}
\author{Yunbo Zhang$^{3}$}
\email{ybzhang@zstu.edu.cn}
\author{Han Pu$^{4}$}
\email{hpu@rice.edu}

\affiliation{
$^1${Institute of Theoretical Physics and State Key Laboratory of Quantum Optics and Quantum Optics Devices, Shanxi University, Taiyuan 030006, China}\\
$^2${Institute for Advanced Study, Tsinghua University, Beijing 100084, China}\\
$^3${Key Laboratory of Optical Field Manipulation of Zhejiang Province and Physics Department of Zhejiang Sci-Tech University, Hangzhou 310018, China}\\
$^4${Department of Physics and Astronomy, and Rice Center for Quantum
    Materials, Rice University, Houston, TX 77005, USA}
}

\maketitle

In this Supplemental Materials (SM), we provide additional information of this work. First, we provide a detailed derivation of the single-particle Hamiltonian, and then explain the symmetry and degeneracy of the single-particle spectrum at $q=0$. Additionally, we discuss the SU(3) operators and the classification of the SU(2) subspaces, and then we show the relations between the Cartesian states and the spin and nematic densities. Furthermore, we specifically display the spin and spin-nematic orders in our single-particle and many-body phase diagrams, and show how various phase transition shown in the main text are classified. Finally, we compare the main results obtained by effective 2D calculation with those obtained by 3D numerical simulations.

\section{Derivation of the Single-Particle Hamiltonian}

\begin{figure}[t]
\includegraphics[width=0.8\textwidth]{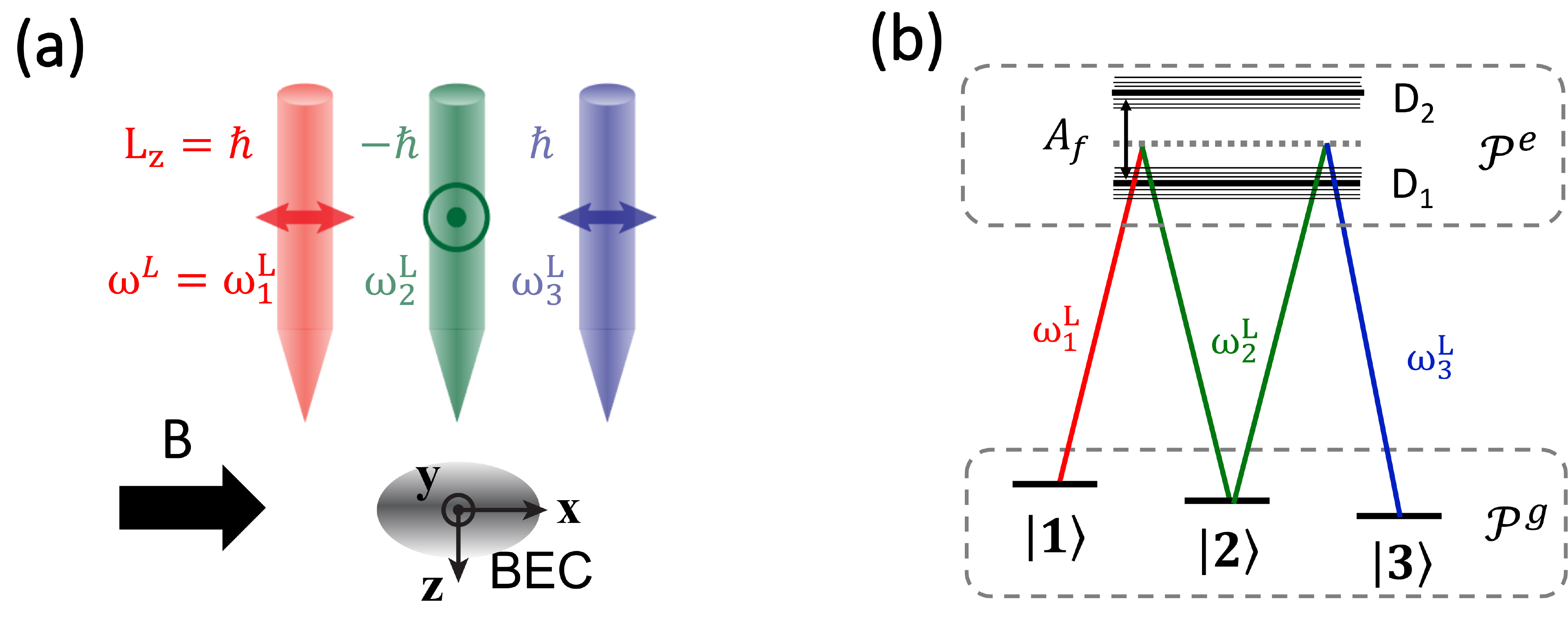}
\caption{Detailed information on the atom-light interactions of our model. {\color{cmagenta}(a) Raman beams with optical frequencies $\omega^L_{j=1,2,3}$ and polarization on directions $\{\hat{e}_x,\hat{e}_y,\hat{e}_x\}$.} Bias magnetic field is fixed along the $x$ direction. (b) Details of atomic level structures, where $\mathcal{P}^g$ and $\mathcal{P}^e$ denote the atomic ground-state and excited-state manifold. D$_1$ and D$_2$ are D-lines with electron angular quantum number $l = 1$, and $A_f$ indicates the fine structure splitting. States $|1\rangle$, $|2 \rangle$ and $|3\rangle$ are the bare atomic spin states defined by the bias field.}
\label{FigS1}
\end{figure}

We provide detailed information of our proposal in Fig.~\ref{S1} as a supplement of the schematic shown in the main text, where three Laguerre-Gaussian laser beams, with optical frequency $\omega^L_{j=1,2,3}$ and orbit angular momentum (OAM) $\hbar$, $-\hbar$ and $\hbar$, respectively, propagate along the $z$-direction and shine on a quasi-2D Bose-Einstein condensate (BEC). A bias magnetic field on the $x$-direction provides a fixed quantization axis. The ground-state $^2$S$_{1/2}$ and the excited manifold $^2$P$_{1/2}$ (D$_1$ line) and $^2$P$_{3/2}$ (D$_2$ line) are coupled by the Raman beams in the way as is shown in Fig.~\ref{S1}(b). Then, the single-particle Hamiltonian reads (we set $\hbar=m=\omega=1$ with $m$ the atomic mass, and $\omega$ the transverse trap frequency)
{\color{cmagenta}
\begin{equation}
\begin{aligned}
H_0 &= h_0 + h_\text{at} + h_\text{int} \\
& = -\frac{1}{2}\nabla ^{2}+\frac{1}{2}r^{2}+\frac{1}{2}\gamma^2z^{2} + \sum_{j \in g}{\omega^g_{j} \mathcal{P}^g_j} + \sum_{j \in e}{\omega^e_{j} \mathcal{P}^e_j} +  A_f \mathbf{L}\cdot \mathbf{S}  -\mathbf{d}\cdot \mathbf{E},
\label{S1}
\end{aligned}
\tag{S1}
\end{equation}
}where $h_0$, consisting of the first five terms on the second line, is the bare Hamiltonian of the single atom with $\mathcal{P}^{g,e} = \sum_{j\in g,e} \mathcal{P}^{g,e}_j = \sum_{j\in g,e} \left|j\right\rangle \left\langle j \right|$ being the projection operators of the ground-state ($g$) or the excited-state ($e$) atomic manifold, and accordingly the $\omega^{g,e}_j$ being the energy of the ground or excited states. The term $h_\text{at} = A_f \mathbf{L}\cdot \mathbf{S}$ represents the fine structure spin-orbit coupling of the valence electron with $A_f$ being the fine-structure interaction strength, and the last term $h_\text{int}=-\mathbf{d}\cdot \mathbf{E}$ represents the atom-light electric dipole interaction with $\mathbf{E}$ the electromagnetic field, and $\mathbf{d}$ the electric dipole moment of the atom.

For the Raman process with large single-photon detuning $\omega^e - \omega^g \gg \omega^L$, the excited states can be adiabatically eliminated by the second-order perturbation theory \cite{Goldman2014} and the resulting effective Hamiltonian in the ground-state manifold is in the form of
\begin{equation}
\begin{aligned}
h_\text{eff} &= -\mathcal{P}^{g} h_\text{int}\mathcal{P}^{e} h_\text{at}^{-1}\mathcal{P}^{e} h_\text{int}\mathcal{P}^{g}\\
 &= \left[u_s |\mathbf{E}|^2 + i u_v (\mathbf{E}^*\times \mathbf{E})\cdot \mathbf{S}\right] \mathcal{P}^{g}.
\label{S2}
\end{aligned}
\tag{S2}
\end{equation}
$h_\text{eff}$ has two main effects. The first term proportional to $u_s\propto \left|\left\langle g|\mathbf{r}|e\right\rangle\right|^2/\Delta$ is the light-induced scalar shift (or ac-Stark shift) being independent on the polarization of the optical beams where $\Delta = \omega^L - (\omega^g-\omega^e)$ is the single-photon detuning; the second term with strength $u_v\propto A_f u_s/\Delta$ is the light-induced vector shift. The scalar shift can be switched off as one chooses the tune-out optical frequency \cite{Schmidt2016} for the Raman beams such that the scalar shifts induced by the D$_2$ and D$_1$ transitions cancel with each other.

Properly engineering the atom-light interaction, the light-induced vector shift would lead to synthetic gauge field or synthetic spin-orbit coupling \cite{Goldman2014}. Particularly for the current Raman configuration in our scheme with $\omega_{1,3}^L$ being linearly polarized along the $x$-direction, and $\omega_{2}^L$ linearly polarized along the $y$-direction, the vector shift induced by the electromagnetic field
\begin{equation}
\begin{aligned}
\mathbf{E} &= \mathbf{E}_1 + \mathbf{E}_2 + \mathbf{E}_3 \\
& =\left[E_1 e^{i(\phi + \omega_1^L t)} + E_3 e^{i(\phi + \omega_3^L t)}\right] \hat{e}_x + E_2 e^{i(-\phi + \omega_2^L t)} \hat{e}_y + c.c.,
\label{S3}
\end{aligned}
\tag{S3}
\end{equation}
can be easily worked out as
\begin{equation}
\begin{aligned}
h_\text{eff} &= \left(\Omega_{12}e^{-i\delta \omega^L_{12} t -i 2i\phi} + \Omega_{23}^*e^{-i\delta \omega^L_{32} t - 2i\phi}\right)S_z + h.c.,
\label{S4}
\end{aligned}
\tag{S4}
\end{equation}
where we have set $\Omega_{ij} = -i u_v {E_i^* E_j}/2$ to be the two-photon Raman frequency and $\delta\omega^L_{ij} = \omega^L_{i}-\omega^L_{j}$, and the $S_z = -(\left| 1 \right\rangle \left \langle 2\right|+ \left| 2 \right\rangle \left \langle 3\right|) + h.c.$ characterizes the particle transitions in the representation of the quantized axis $S_x$. However, in Eq.~(\ref{S4}), not all the transitions are allowed by the level diagram Fig.~\ref{S1}(b). Neglecting the forbidden transitions and the virtual photon processes (counter-rotating-wave terms), and then by taking $\Omega_{12} = \Omega_{23} = \Omega_R$, we have the simplified total Hamiltonian as
{\color{cmagenta}
\begin{equation}
\begin{aligned}
H_0 = -\frac{1}{2}\nabla ^{2}+\frac{1}{2}r^{2}+\frac{1}{2}\gamma^2z^{2} + \sum_{j \in g}{\omega^g_{j} \mathcal{P}^g_j} + \Omega_R
\begin{pmatrix}
	0 & e^{-2i\phi-i\delta\omega_{12}^Lt} & 0\\
	e^{2i\phi+i\delta\omega_{12}^Lt}& 0  & e^{2i\phi+i\delta\omega_{32}^Lt}\\
	0 & e^{-2i\phi-i\delta\omega_{32}^Lt} & 0
\end{pmatrix}.
\label{S5}
\end{aligned}
\tag{S5}
\end{equation}
}Under a standard procedure, we rewrite the Hamiltonian (\ref{S5}) in a rotating frame defined by the unitary operator $U=e^{i(\delta\omega_{12}^L \mathcal{P}^g_1 + \delta\omega_{32}^L\mathcal{P}^g_{3})t}$ as
{\color{cmagenta}
\begin{equation}
\begin{aligned}
H_0 = -\frac{1}{2}\nabla ^{2}+\frac{1}{2}r^{2}+\frac{1}{2}\gamma^2z^{2} +
\begin{pmatrix}
	q + \delta & 0 & 0\\
	0 & 0  & 0\\
	0 & 0 & q - \delta
\end{pmatrix}
+ \Omega_R
\begin{pmatrix}
	0 & e^{-2i\phi} & 0\\
	e^{2i\phi}& 0  & e^{2i\phi}\\
	0 & e^{-2i\phi} & 0
\end{pmatrix},
\label{S6}
\end{aligned}
\tag{S6}
\end{equation}
}where we have used the relations $\delta = (\delta_{32} - \delta_{12})/2$ and $q = (\delta_{32} + \delta_{12})/2$ with $\delta_{ij} = \delta\omega^L_{ij} - (\omega_{i}^g-\omega_{j}^g)$ being the two-photon detuning. Finally, a global spin rotation $S_x\rightarrow S_z$, $S_z\rightarrow S_y$ and $S_y\rightarrow S_x$ helps us to enter the commonly used $S_z$ representation and transforms Hamiltonian (\ref{S6}) into Hamiltonian (1) in the main text.

Here, we would like to make two comments. At first, the above discussion is general and can be in principle applied to any alkaline-metal atomic species in BEC experiments. Let us take the $^{87}$Rb atom as a specific example, which has been widely used in the experiments involving spin-orbit coupled BEC \cite{Lin2011,HChen2018,PChen2018}. For $^{87}$Rb atom, one can choose the ground-state Zeeman levels  $5^2$S$_{1/2}$ $\left| F=1,m_F=\pm1,0\right\rangle$ and the corresponding D-line transitions $5^2$P$_{1/2}$ (D$_1$ line) and $5^2$P$_{3/2}$ (D$_2$ line) to construct the level configuration shown in the Fig.~\ref{S1}(b), and take $\sim 790$nm as the wave length of the Raman beams to match the tune-out wave length of $^{87}$Rb \cite{Schmidt2016}. Furthermore, the condition $\Omega_{12} = \Omega_{23} = \Omega_R$ can be realized since the strength of $|E_{i=1,2,3}|\propto \sqrt{I_i}$ of the three Raman beams can be tuned independently, with $I_i$ the light intensity of the Raman beams. {\color{cmagenta}Secondly, note that the transverse coordinates $(r,\phi)$ and the longitudinal coordinate $z$ are decoupled in Eq.~(\ref{S6}), and the Raman-induced spin-orbital coupling (last term of Eq.~(\ref{S6})) only lies in the transverse plane, which thus allow us to make the on-going calculations solely in the $(r,\phi)$ 2D system as were shown in the main text. In the last section of this SM, we carry out a fully 3D numerics by solving 3D Gross-Pitaevskii equations, and show the 3D results are highly consistent with those obtained by the 2D calculation.}

\section{Spectrum At Zero Quadratic Zeeman Splitting}
An apparent feature in the case of $q=0$ is that the energy spectrum of the even-parity states are symmetric about $l_z=1$. It means there is a symmetry only existing in the even-parity subspace of the Hamiltonian $\tilde{H}_{0}$ (Eq.~(2) in the main text). The even-parity subspace is spanned by the even-parity Cartesian states $\left|y\right\rangle$ and $\left|z\right\rangle$, under which basis, $\tilde{H}_{0}$ can be written into the matrix form
\begin{equation}
\tilde{H}_{0} = -\frac{\partial^2}{2\partial r^2}-\frac{\partial}{2r\partial r}+\frac{r^{2}}{2}-\frac{1}{2r^2}
\begin{bmatrix}
	(l_z-2)^2 & 0\\
	0 & l_z^2
\end{bmatrix}
+\sqrt{2}\Omega _{R}
\begin{bmatrix}
	0 & -i\\
	i & 0
\end{bmatrix}
,
\label{S7}
\tag{S7}
\end{equation}
where we have used the relations $\nabla^2 = \partial_r^2 + \partial_r /r + \partial^2/r^2\partial\phi^2$ and $L_z = -i\partial_\phi$, and the properties of the Cartesian states that we will discuss in details in the following section.

Considering $l_z = 1 + l_z'$, we have
\begin{equation}
\tilde{H}_{0} = -\frac{\partial^2}{2\partial r^2}-\frac{\partial}{2r\partial r}+\frac{r^{2}}{2}-\frac{1}{2r^2}
\begin{bmatrix}
	(l_z'-1)^2 & 0\\
	0 & (l_z'+1)^2
\end{bmatrix}
+\sqrt{2}\Omega _{R}
\begin{bmatrix}
	0 & -i\\
	i & 0
\end{bmatrix}
,
\label{S8}
\tag{S8}
\end{equation}
and $l_z = 1 - l_z'$ which leads to
\begin{equation}
\tilde{H}_{0} = -\frac{\partial^2}{2\partial r^2}-\frac{\partial}{2r\partial r}+\frac{r^{2}}{2}-\frac{1}{2r^2}
\begin{bmatrix}
	(l_z'+1)^2 & 0\\
	0 & (l_z'-1)^2
\end{bmatrix}
+\sqrt{2}\Omega _{R}
\begin{bmatrix}
	0 & -i\\
	i & 0
\end{bmatrix}
.
\label{S9}
\tag{S9}
\end{equation}
The two Hamiltonians (\ref{S8}) and (\ref{S9}) should have the same spectrum as they are related by a unitary transformation
\begin{equation}
\mathcal{U} =
\begin{bmatrix}
	0 & i\\
	-i & 0
\end{bmatrix}
.
\label{S10}
\tag{S10}
\end{equation}
As a result, given a state with quasi-OAM $(1+ l_z' )$, there exists a degenerate state with quasi-OAM $(1- l_z' )$. Hence the spectrum is symmetric about $l_z=1$.

\section{SU(3) Operators and Subspaces Classification}
In the main text, we defined a series of SU(3) operators including the three spin operators $S_{\mu=x,y,z}$ and nine symmetrized nematic operators $N_{\mu\nu} = \frac{1}{2}\left(S_\mu S_\nu+S_\nu S_\mu\right)-\frac{2}{3}\delta_{\mu\nu}$. Under the basis of bare spin states $\left|\pm1\right\rangle$ and $\left|0\right\rangle$, these operators have the following explicit matrix form:
\begin{equation}
\begin{aligned}
S_x &=
\frac{1}{\sqrt{2}}\begin{pmatrix}
0 & 1 & 0 \\
1 & 0 & 1 \\
0 & 1 & 0%
\end{pmatrix}, \ \ \
S_y = \frac{i}{\sqrt{2}}
\begin{pmatrix}
0 & -1 & 0 \\
1 & 0 & -1 \\
0 & 1 & 0%
\end{pmatrix}, \ \ \
S_z =
\begin{pmatrix}
1 & 0 & 0 \\
0 & 0 & 0 \\
0 & 0 & -1%
\end{pmatrix},\\
N_{xx} &=
\frac{1}{6}
\begin{pmatrix}
-1 & 0 & 3 \\
0 & 2 & 0 \\
3 & 0 & -1%
\end{pmatrix},  \ \ \
N_{yy} =
\frac{1}{6}
\begin{pmatrix}
-1 & 0 & -3 \\
0 & 2 & 0 \\
-3 & 0 & -1%
\end{pmatrix},  \ \ \
N_{zz} =
\frac{1}{3}
\begin{pmatrix}
1 & 0 & 0 \\
0 & -2 & 0 \\
0 & 0 & 1%
\end{pmatrix},\\
N_{xy} &=
\frac{1}{2}
\begin{pmatrix}
0 & 0 & i \\
0 & 0 & 0 \\
-i & 0 & 0%
\end{pmatrix},  \ \ \
N_{yz} = \frac{i}{2\sqrt{2}}
\begin{pmatrix}
0 & -1 & 0 \\
1 & 0 & 1 \\
0 & -1 & 0%
\end{pmatrix},  \ \ \
N_{zx} =
\frac{1}{2\sqrt{2}}
\begin{pmatrix}
0 & 1 & 0 \\
1 & 0 & -1 \\
0 & -1 & 0%
\end{pmatrix},
\label{S11}
\end{aligned}
\tag{S11}
\end{equation}
with $N_{yx}=N_{xy}$, $N_{zy}=N_{yz}$, and $N_{xz}=N_{zx}$ by definition. However, only eight of the above operators are linearly independent, and form a complete set of generators of the SU(3) Lie group laying as the mathematical foundation of the spin-1 quantum system.

The SU(3) group has a large number of SU(2) subgroups (or SU(2) subspaces) which are generated by triads of operators $\{\hat{O}_i,\hat{O}_j,\hat{O}_k\}$ satisfying cyclic commutation relation $[\hat{O}_i,\hat{O}_j] = i\alpha\epsilon_{ijk}\hat{O}_k$ where $\alpha$ is the structure constant and $\epsilon_{ijk}$ is the Levi-Civita antisymmetric tensor. Root diagram obtained in the adjoint representation of the Cartan subalgebra provides a powerful way in identifying all the SU(2) subspaces \cite{Yukawa2013}. The subspaces can and can only be classified into two types with structure constant $\alpha$ being equal to 1 and 2, respectively. The most typical type-1 subspace that has received tremendous attention is the spin subspace $\mathbf{S} = \{S_x,S_y,S_z\}$ with $\alpha=1$. The spin-nematic subspace $\mathbf{Q}=\{2N_{yz},S_x,D_{yz}=N_{yy}-N_{zz}\}$ used in the main text is of type-2 with $\alpha=2$. It has been proven that the SU(3) rotations can transform subspaces belonging to the same type, but cannot transform those belonging to different types \cite{Yukawa2013}. Since $\mathbf{Q}$ can generate a SU(2) group being isomorphic to the three-dimensional rotational group SO(3), we can treat $\mathbf{Q}$ as a vector operator. Consequently, in the space of $\mathbf{Q}$, an arbitrary three-dimensional rotation is characterized by an unitary transformation $U=e^{-i\frac{\phi}{2} \mathbf{n}\cdot\mathbf{Q}}$ with $\mathbf{n} = (n_x,n_y,n_z)^T$ and $\phi$ corresponding to the rotational axis and the rotational angle respectively, and the factor $1/2$ in the exponent appeared due to the structure constant being $\alpha=2$.

\section{Cartesian States and Spin-Nematic Density}
As is shown in the main text, the Cartesian states $\left|\mu = x,y,z\right\rangle$ are eigenstates of the spin operators $S_\mu$ with zero eigenvalues, i.e. $S_\mu \left|\mu\right\rangle=0$ \cite{Ohmi1998}. The transformation matrix between the bare states and the Cartesian states is given by
\begin{equation}
U =
\begin{pmatrix}
-\frac{1}{\sqrt{2}} & \frac{i}{\sqrt{2}} & 0 \\
0 & 0 & 1 \\
\frac{1}{\sqrt{2}} & \frac{i}{\sqrt{2}} & 0
\end{pmatrix}.
\label{S12}
\tag{S12}
\end{equation}
In the Cartesian basis, the spin and nematic operators discussed above are in quite simple forms of
\begin{equation}
\left\langle \mu | S_\eta | \nu \right\rangle = i \epsilon_{\mu\eta \nu},
\label{S13}
\tag{S13}
\end{equation}
and
\begin{equation}
\left\langle \mu | N_{\eta\gamma} | \nu \right\rangle = -\frac{1}{2}(\delta_{\mu\eta}\delta_{\gamma\nu}+\delta_{\mu\gamma}\delta_{\eta\nu}) + \frac{1}{3}\delta_{\mu\nu}\delta_{\eta\gamma},
\label{S14}
\tag{S14}
\end{equation}
where subscript labels $\{\mu,\nu,\eta,\gamma\}$ can take $\{x,y,z\}$.

Consider an arbitrary state expanded using the Cartesian basis $\left|\psi\right\rangle = \sum_{\mu} m_\mu \left| \mu \right\rangle$, the expectation of the spin operators are
\begin{equation}
\begin{aligned}
\mathcal{S}_\eta &= \left\langle \psi | S_\eta| \psi \right\rangle \\
& =\sum_{\mu,\nu} m^*_\mu m_\nu \left\langle \mu | S_\eta| \nu \right\rangle \\
& =i\sum_{\mu,\nu} m^*_\mu m_\nu \epsilon_{\mu\eta \nu}, \\
\label{S15}
\end{aligned}
\tag{S15}
\end{equation}
or more compactly:
\begin{equation}
\boldsymbol{\mathcal{S}} = -i \mathbf{m}^* \times \mathbf{m};
\label{S16}
\tag{S16}
\end{equation}
the expectation of the spin nematic operators are
\begin{equation}
\begin{aligned}
\mathcal{N}_{\eta\gamma} &= \left\langle \psi | N_{\eta\gamma}| \psi \right\rangle \\
& =\sum_{\mu,\nu} m^*_\mu m_\nu \left[-\frac{1}{2}(\delta_{\mu\eta}\delta_{\gamma\nu}+\delta_{\mu\gamma}\delta_{\eta\nu}) + \frac{1}{3}\delta_{\mu\nu}\delta_{\eta\gamma}\right] \\
& =\frac{\delta_{\eta\gamma}}{3}-\frac{1}{2}\sum_{\mu,\nu} m^*_\mu m_\nu (\delta_{\mu\eta}\delta_{\gamma\nu}+\delta_{\mu\gamma}\delta_{\eta\nu}), \\
& =\frac{\delta_{\eta\gamma}}{3}-\frac{1}{2}\left(m^*_\eta m_\gamma+m^*_\gamma m_\eta\right),
\label{S17}
\end{aligned}
\tag{S17}
\end{equation}
or more compactly
\begin{equation}
\mathcal{N} = \frac{1}{3} - \text{Re}[\mathbf{m}^*\otimes\mathbf{m}],
\label{S18}
\tag{S18}
\end{equation}
where $\otimes$ denotes the Kronecker product. With Eqs.~(\ref{S16}) and (\ref{S18}), one can easily obtain the spin and nematic densities shown in the main text.

\section{Quantum Phase Transitions}

\begin{figure}[t]
\includegraphics[width=0.8\textwidth]{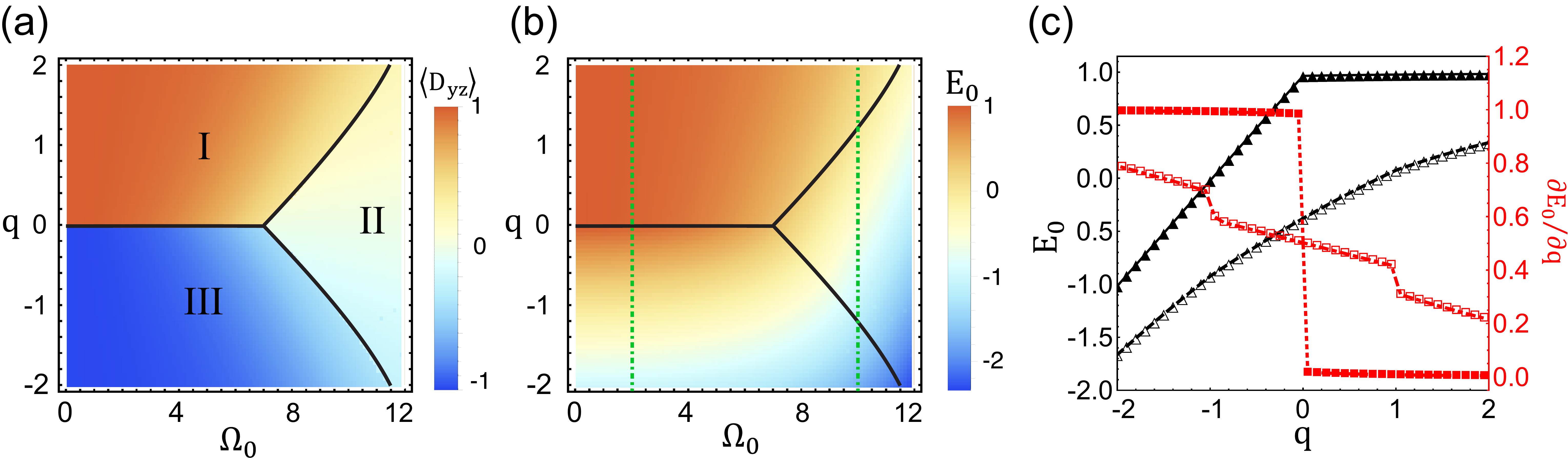}
\caption{Single-particle phase diagrams and phase transitions. (a) Dependence of the averaged longitudinal nematic order $\left\langle \mathcal{D}_{yz}\right\rangle$ on $\Omega_0$ and $q$. (b) Dependence of the ground-state energy $E_0$ on $\Omega_0$ and $q$. (c) Variation of the ground-state energy $E_0$ (solid and dashed lines with triangles) and the its first derivative $\partial_q E_0$ (dotted and dot-dashed lines with squares) at fixed $\Omega_0 = 2$ and $\Omega_0 = 10$, as are indicated by the two green dot-dashed lines in (b), where $\partial_q E_0$ shows discontinuity at the first-order phase transition point. }
\label{FigS2}
\end{figure}

In the main text, we show two phase diagrams (single-particle and many-body phases diagrams), and various quantum phase transitions that can be either first-order or second-order. Here, we present detailed information on the phase diagrams and the classification of the phase transitions.

Considering that we are dealing with the case at zero temperature $T=0$, the ground-state energy $E_0$ is the quantity that we are mainly interested in. Besides, since the quantum state $\boldsymbol{\Psi}$ carries both spin and spin-nematic orders, the  averaged spin $\left\langle S_\mu \right\rangle$ and nematicity $\left\langle N_{\mu\nu} \right\rangle$ serve as macroscopic order parameters that can be used in phase identification. The averaged spin and nematicity are defined as the spatial average of the local ones, i.e.
\begin{equation}
\left\langle S_\mu \right\rangle = \int d^2\mathbf{r} \boldsymbol{\Psi}^\dagger S_\mu \boldsymbol{\Psi} = \int d^2\mathbf{r} \rho(\mathbf{r}) \mathcal{S}_\mu(\mathbf{r}),
\label{S19}
\tag{S19}
\end{equation}
and
\begin{equation}
\left\langle N_{\mu\nu} \right\rangle = \int d^2\mathbf{r} \boldsymbol{\Psi}^\dagger N_{\mu\nu} \boldsymbol{\Psi} = \int d^2\mathbf{r} \rho(\mathbf{r}) \mathcal{N}_{\mu\nu}(\mathbf{r}),
\label{S20}
\tag{S20}
\end{equation}
where $\mathcal{S}_\mu$ and $\mathcal{N}_{\mu\nu}$ are the normalized spin and nematic densities defined in Eqs.~(6) and (7) in the main text.

In general, a first-order phase transition is featured by a discontinuity of the first-order derivative of $E_0$, and at the same time the order parameter exhibits a sudden jump; whereas a second-order transition is continuous in the first-order derivative of $E_0$, but is discontinuous in the second-order derivative, and in the meanwhile, the order parameter goes smoothly from a finite value to zero or vice verse. Therefore, the behaviors $E_0$ and the averaged spin/spin-nematic orders help us distinguish different phases as well as the order of the phase transition.

For the single-particle phase diagram shown in Fig.~2(a) in the main text, the phases I, II and III feature vanishing total spins $\left|\left\langle \mathbf{S}\right\rangle\right|$ but non-vanishing averaged nematic order. In Figs.~\ref{S2}(a) and (b), we reprint the single-particle phase diagram with background colors showing the variations of the longitudinal nematic order $\left\langle \mathcal{D}_{yz}\right\rangle$ and the ground-state energy $E_0$, respectively. One can observe that the nematic order $\left\langle \mathcal{D}_{yz}\right\rangle$ exhibits sudden jumps across the phase boundaries indicating all the transitions among phases I, II and III are of first order (labeled by solid lines). The order of the phase  transitions are confirmed by examining the behavior of $E_0$ and $\partial_q E_0$ as functions of $q$, as shown in Fig.~\ref{S2}(c). In Fig.~\ref{S2}(c), the lines with solid markers are plotted in the case of $\Omega_0=2$ where the transition I-III occurs at $q=0$, and the lines with hollow markers are plotted in the case of $\Omega_0=10$ where two transitions III-II and II-I occur at $q\approx\pm1$, respectively. The two cases are visually indicated by the two vertical dot-dashed lines in Fig.~\ref{S2}(b). It is clearly shown in the Fig.~\ref{S2}(c) that the energy $E_0$ varies continuously, but its first-order derivative $\partial_q E_0$ shows discontinuous jumps at the phase boundaries.

Similar analyses are performed on the many-body phase diagram displayed in Fig.~\ref{S3}, where subfigures (a)-(c) show the dependence of $\left\langle \mathcal{D}_{yz}\right\rangle$, the total spin $\left|\left\langle \mathbf{S}\right\rangle\right| = \sqrt{\left\langle S_x \right\rangle^2 + \left\langle S_y \right\rangle^2+ \left\langle S_z \right\rangle^2}$, and $E_0$ on the phase plane $\Omega_0$-$q$. The two emergent new phases IV and V are ferromagnetic with non-vanishing total spin, i.e. $\left|\left\langle \mathbf{S}\right\rangle\right| \neq0$. As mentioned in the main text, these two new phases break different symmetries  (Phase IV keeps the rotational symmetry but breaks the spin-parity symmetry; whereas Phase V keeps the spin-parity symmetry but breaks the rotational symmetry), and hence the phase transition between them is of first-order (denoted by solid lines), as confirmed by a sudden jump of the first-order derivative $\partial_qE_0$ across the phase boundary (not shown in the Fig.~\ref{S3}). In contrast, the phase transitions between the symmetry preserved phases (I, II and III) and the symmetry broken phases (IV and V) are all of the second order (denoted by dashed lines). We examine these transitions by tracking the variational amplitudes $|D_{0,1,+,-}|$ (upper panel of Fig.~\ref{S3}(e)), the averaged spin and spin-nematic orders (lower panel of Fig.~\ref{S3}(e)) and the energy behaviors (Fig.~\ref{S3}(f)) at $\Omega_0=2.5$ (indicated by the vertical dot-dashed line in Fig.~\ref{S3}(c)), where three transitions IV-III, III-V, and V-I occur as $q$ is ascendingly swept. The second-order transitions are clearly demonstrated in Fig.~\ref{S3}(f) as the second-order derivative $\partial_q^2E_0$ is the lowest order of derivatives that exhibits a discontinuity at the phase boundary, and at the same time the total spin order $\left|\left\langle \mathbf{S}\right\rangle\right|$ varies smoothly from a finite value to zero or from zero to a finite value. Additionally, we note that,  in the limit of $\Omega_0\rightarrow0$ (namely the Raman lights being switched off), the phases IV, V and I in the diagram Fig.~\ref{S3}(b) will reduce to the conventional ferromagnetic phase, the broken-axisymmetry phase and the polar phase of a spin-1 BEC without spin-orbit coupling \cite{Stenger1998,Kawaguchi2012}. These three conventional phases exist in the regimes $q<0$, $q\gtrsim0$ and $q\gg0$, and are featured by fully longitudinal magnetization $\left|\left\langle S_z \right\rangle\right|=1$, finite transverse magnetization $\left|\left\langle S_\perp \right\rangle\right|\neq 0$ and vanishing magnetization $\left|\left\langle \mathbf{S}\right\rangle\right|=0$, respectively.

\begin{figure}[t]
\includegraphics[width=0.8\textwidth]{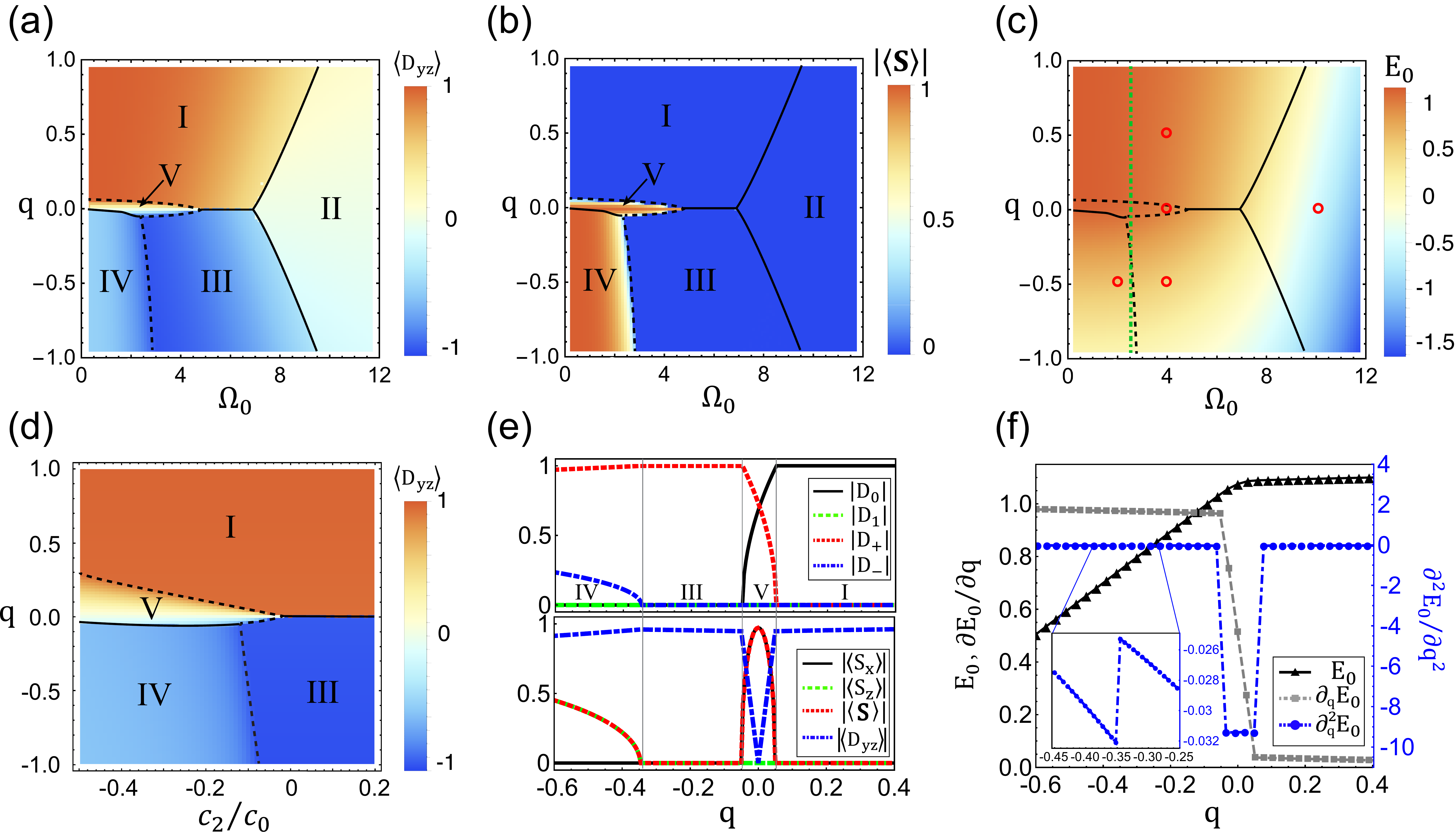}
\caption{Many-body phase diagrams and phase transitions. (a)-(c) Many-body phase diagrams on the $\Omega_0$-$q$ plane with fixed $c_2/c_0=-0.1$, where the background colors in (a), (b) and (c) denote the averaged longitudinal nematic order $\left\langle \mathcal{D}_{yz}\right\rangle$, total spin order $\left|\left\langle \mathbf{S}\right\rangle\right| = \left[\left\langle S_x \right\rangle^2 + \left\langle S_y \right\rangle^2+ \left\langle S_z \right\rangle^2\right]^{1/2}$ and the ground-state energy $E_0$, respectively. {\color{cmagenta}Red circles in subfigure (c) indicate the typical points where the three-dimensional ground states are shown in Fig.~\ref{S4}.}
(d) Many-body phase diagram and dependence of the $\left\langle \mathcal{D}_{yz}\right\rangle$ on the $q$-$c_2/c_0$ plane, where $\Omega_0 = 2.5$ is fixed.
In all the phase diagrams (a)-(d), the black solid and the black dashed lines indicate the first- and the second-order phase boundaries, respectively.
(e) Variations of the variational amplitudes $|D|$, and the total spin and nematic orders on $q$. Upper panel: dependence of the variational amplitudes $|D_{0,1,+,-}|$. Lower panel: dependence of the total spin and spin-nematicity $|\left\langle S_x \right\rangle|$, $|\left\langle S_z \right\rangle|$, $\left|\left\langle \mathbf{S}\right\rangle\right|$, and $|\left\langle D_{yz} \right\rangle|$.
(f) Ground-state energy $E_0$ and its first $\partial_q E_0$ and second derivatives $\partial_q^2 E_0$, where $\partial_q E_0$ and $\partial^2_q E_0$ exhibit discontinuity at the first- and the second-order phase boundaries, respectively. Insets: a closed look at the $\partial^2_q E_0$ in the regime $q\in[-0.45,-0.25]$. In subfigures (e) and (f), we take $\Omega_0 = 2.5$ and $c_2=-0.1c_0$ as is indicated by the green dot-dashed line in subfigure (c).}
\label{FigS3}
\end{figure}

In the above we have shown that the weak many-body interaction would lead to new phases that do not appear in the single-particle picture. The emergence of the new phases can be attributed to the interplay between the single-particle Hamiltonian $H_0$, the density-density interaction $c_0$ term and the spin-dependent interaction $c_2$ term in Eq.(13) of the main text. Here, let us make this point more clear by re-examining each term in the Hamiltonian. The single-particle Hamiltonian carries both rotational and spin-parity symmetry as mentioned before, which renders the single particle states to be both rotational symmetric and total-spin vanished since the local spin density $\boldsymbol{\mathcal{S}}$ (Eq.(8) in the main text) cancel with each other during the spatial average (see Fig.~\ref{S3}(b) and Eq.~(\ref{S19})). {\color{cGreen}The density-density interaction $c_0>0$, which is SU(3) symmetric, only tends the total density distribution $\rho(\mathbf{r})$ of the condensate to be more uniform and extended in order to minimize the density-density interaction energy. Hence, even in the case of $q=0$ where the two vortex states ($\boldsymbol{\Psi}_{+,0}$ and $\boldsymbol{\Psi}_{+,2}$) are energy degenerate, the $c_0$ term tends the ground state of the condensate to be one of these two states, rather than their superposition, since the superposition would break the rotational symmetry and thus make the density distribution $\rho(\mathbf{r})$ to be more localized (hence increasing interaction energy). Therefore, from this point of view, the rotational symmetry as well as the corresponding spin-nematic topology of the two vortices ($\boldsymbol{\Psi}_{+,0}$ and $\boldsymbol{\Psi}_{+,2}$) are 'protected' by the density-density interaction $c_0$.}

In contrast to $c_0$, the spin-dependent $c_2$ term is highly magnetization-related. This term favors anti-ferromagnetic states with vanishing macroscopic polarization if $c_2>0$, which is also called the anti-ferromagnetic interaction; whereas favors ferromagnetic states if $c_2<0$, which is thus called the ferromagnetic interaction. Consequently, the single-particle states in phase I, II and III are more preferred by $c_2\ge0$ since there is no obvious confliction among the three terms ($H_0$, $c_0$, and $c_2$ terms) mentioned above. However if $c_2<0$, the $c_2$ term would strongly compete with the other two, since generating a state with macroscopic magnetization would either break the spin-parity symmetry which is not favored by $H_0$, or break the rotational symmetry which is not preferred by either $H_0$ or $c_0$ term. This competition consequently leads to the emergent new phases IV and V whose areas are highly dependent on the strength of $c_2$. In Fig.~\ref{FigS3}(d), we display the phase dependence on $c_2$ by fixing $\Omega_0=2.5$ and $c_0=1$. It is clearly shown that, the emergent phases IV and V only exist in the region of $c_2<0$, and the corresponding phase areas are significantly squeezed as $|c_2|$ decreases to zero, in accordance with the qualitative analysis above. We additionally note that, for the most commonly used alkaline metal with ferromagnetic interaction in experiments, $^{87}$Rb for example, the intrinsic spin-exchange interaction $|c_2/c_0|$ is about $0.5\%$ \cite{Lin2011}, which is thus too weak to generate a considerable observation effect especially for the angular striped phase V. One possible solutions is to adopt the optical Feshbach resonance where $c_2$ is supposed to be enlarged by one order of magnitude \cite{Hamley2008}. However, the life time of the BEC will also be suppressed due to the optical heating. Another more feasible solution may be to model on pseudo-spins defined on multiple wells or optical lattice. This method has already been implemented on a spin-1/2 BEC with SOC to observe the striped phase \cite{Li2017}, where the spin-dependent interaction can be adjusted in an easier way.

{\color{cmagenta}\section{Three-Dimensional Simulation and the Gross-Pitaevskii Equations}}
\begin{figure}[t]
\includegraphics[width=0.8\textwidth]{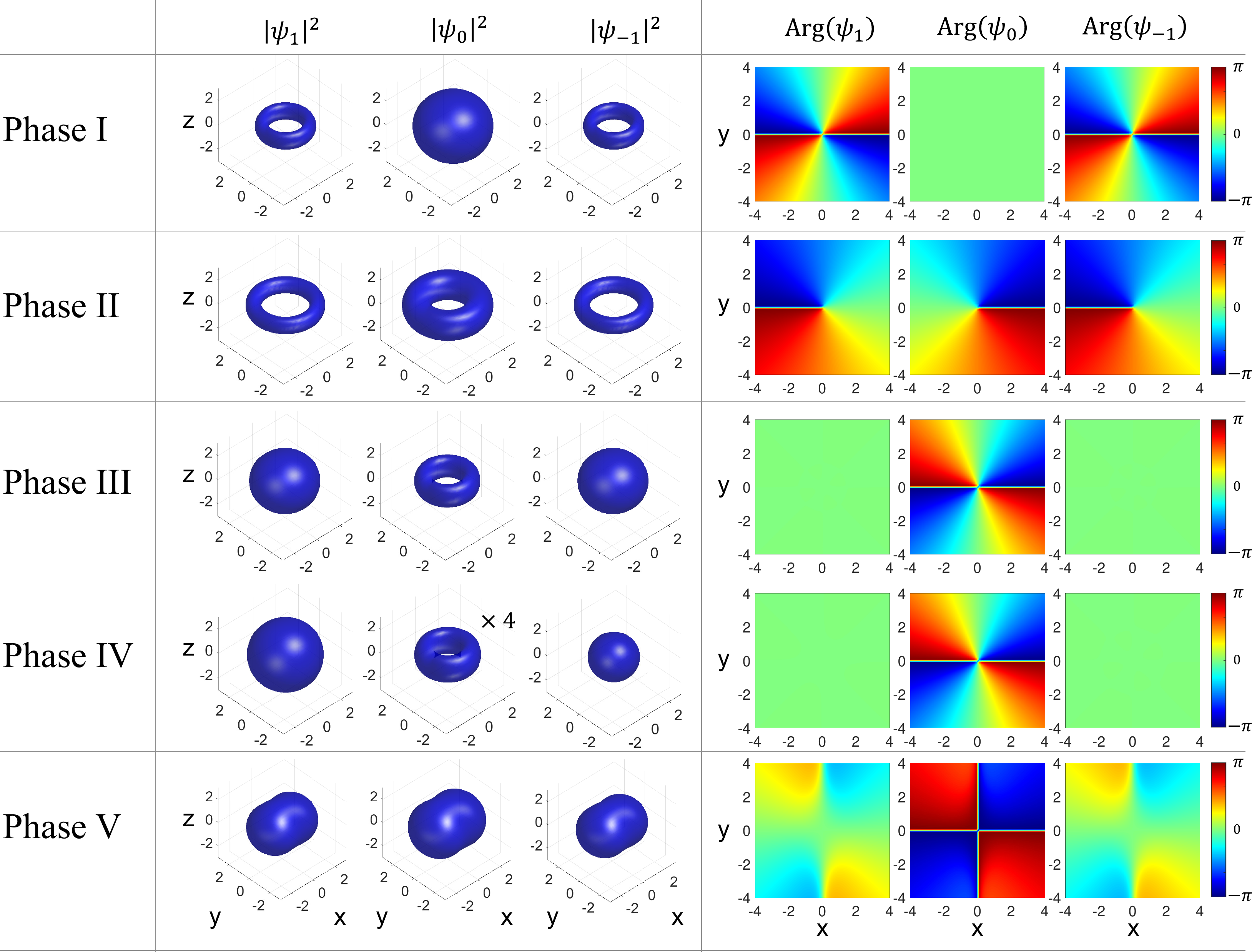}
\caption{{\color{cmagenta}Density profiles and phase distributions of the three-dimensional BEC in the lab frame. Left panel: isosurface plot of the density distribution $|\psi(\mathbf{r},z)|^2$ in each phases, where the 1st, 2nd and 3rd column corresponds to the density distribution on the bare spin components 1, 0, and -1, respectively. Particularly for Phase IV, we enlarge the 0-component density $\psi_0$ by four times such that the density distribution can be clearly displayed. Right panel: the corresponding phase distribution of the many-body wave function in the x-y plane with $z=0$, i.e. $\text{Arg}(\psi_i(\mathbf{r},z=0))$. In our calculation, we considered an isotropic 3D harmonic confinement by taking $\gamma = \omega_z/\omega = 1$, and $c^\text{3D}_0 = \sqrt{2\pi}$ and $c_2^\text{3D} = -0.1 c_0^{\text{3D}}$. The specific $\Omega_0$ and $q$ we take in each phase are: $\{\Omega_0,q\} = \{4,0.5\}_\text{I}$, $\{10,0\}_\text{II}$, $\{4,-0.5\}_\text{III}$, $\{2,-0.5\}_\text{IV}$, $\{4,0\}_\text{V}$, as visually labeled out by red circles in Fig.~\ref{FigS3}(c).}}
\label{FigS4}
\end{figure}

{\color{cmagenta}Up to now, all the discussions are performed in the two-dimensional $x$-$y$ plane as mentioned in the modeling section of the main text. However, considering the recent two spin-OAM experiments \cite{HChen2018,PChen2018,Zhang2018} were done in three dimensional BEC systems, therefore we are motivated to perform a fully three-dimensional numerics by solving the coupled Gross-Pitaevskii(GP) equations and show the main physics obtained above remains valid in 3D BECs. We derive the GP equations from the total Hamiltonian $H = H_0 + H_\text{int}$ in the lab frame, where $H_0$ and $H_\text{int}$ are single-particle and interacting Hamiltonian shown as Eq.~(1) and Eq.~(13) in the main text. Then the GP equations are explicitly given by
\begin{equation}
\begin{aligned}
i\partial_t \psi_1 &= \left(-\frac{\nabla^2}{2} + \frac{r^2}{2} + \frac{\gamma^2 z^2}{2} + q+\delta + c^\text{3D}_0 \rho+ c^\text{3D}_2 \rho \mathcal{S}_{z} \right )\psi_1 + \left(\Omega_R e^{-2i\phi} + \frac{c^\text{3D}_2}{\sqrt{2}} \rho \mathcal{S}_{-} \right) \psi_0 , \\
i\partial_t \psi_0 &= \left(-\frac{\nabla^2}{2} + \frac{r^2}{2} + \frac{\gamma^2 z^2}{2} + c^\text{3D}_0 \rho\right)\psi_0 +
\left( \Omega_Re^{2i\phi} + \frac{c^\text{3D}_2}{\sqrt{2}} \rho \mathcal{S}_{+}\right) \psi_1
+\left( \Omega_Re^{2i\phi} + \frac{c^\text{3D}_2}{\sqrt{2}} \rho \mathcal{S}_{-}\right) \psi_{-1}, \\
i\partial_t \psi_{-1} &= \left(-\frac{\nabla^2}{2} + \frac{r^2}{2} + \frac{\gamma^2 z^2}{2}+ q - \delta + c^\text{3D}_0 \rho- c^\text{3D}_2 \rho \mathcal{S}_{z} \right )\psi_{-1} + \left(\Omega_R e^{-2i\phi} + \frac{c^\text{3D}_2}{\sqrt{2}} \rho \mathcal{S}_{+} \right) \psi_{0} ,
\end{aligned}
\label{S21}
\tag{S21}
\end{equation}
where $\rho(\mathbf{r},z) = \left| \boldsymbol{\Psi}(\mathbf{r},z) \right|^2$ is the 3D total density,
\begin{equation}
	\mathcal{S}_{\mu=x,y,z}(\mathbf{r},z)=\frac{\mathbf{\Psi}^\dagger(\mathbf{r},z) S_\mu \mathbf{\Psi}(\mathbf{r},z)}{\rho(\mathbf{r},z)}
	\label{S22}
	\tag{S22}
\end{equation}
 is the 3D normalized spin density with $\mathcal{S}_\pm = \mathcal{S}_x \pm i \mathcal{S}_y$, and $c^\text{3D}_{0,2}$ are the 3D interaction strengths, which are related to the 2D ones by $c^\text{3D}_{0,2} = \sqrt{2\pi/\gamma}c_{0,2}$ to a good approximation \cite{Bao2013} with $\gamma = \omega_z/\omega$ being the aspect ratio of the confinements.
With equation Eq.~(\ref{S21}), we obtain the many-body ground states $\mathbf{\Psi}(\mathbf{r},z)$ by propagating the GP equations in imaginary time, which is commonly called the imaginary-time evolution. In our numerics, we deal with the temporal propagation using the time-splitting method \cite{Bao2013}, and employ the pseudo-spectral method and the finite difference method to deal with the kinetic term $-\nabla^2/2$ and the other non-kinetic terms, respectively.

In our GP simulation, we confine the BEC confined in a 3D isotropic harmonic trap by taking $\gamma = 1$, $c^\text{3D}_0 = \sqrt{2\pi}$ and $c_2^\text{3D} = -0.1 c_0^{\text{3D}}$, and the typical ground-state density distributions as well as the transverse phase windings in each phases are displayed in Fig.~\ref{FigS4}. Specifically, in the left panel of Fig.~\ref{FigS4}, we show the isosurface plots of the density distributions $|\psi_{i=1,0,-1}(\mathbf{r},z)|^2$ of different bare spin components, where the specific $\{\Omega_0,q\}$ used in the calculation are: $\{\Omega_0,q\} = \{4,0.5\}_\text{I}$, $\{10,0\}_\text{II}$, $\{4,-0.5\}_\text{III}$, $\{2,-0.5\}_\text{IV}$, $\{4,0\}_\text{V}$ as visually indicated by red circles in Fig.~\ref{S3}(c); in the right panel of Fig.~\ref{FigS4}, we accordingly plot the transverse phase distributions at $z=0$ plane, i.e. $\text{Arg}(\psi_i(\mathbf{r},z=0))$. One can clearly observe in Fig.~\ref{FigS4} that, the former four phases (I, II, III, and IV) are rotationally symmetric in the $x$-$y$ plane and each spin components carrying quantized phase windings labeled by the red numbers in Fig.~2(b) of the main text. The spin-parity symmetry breaking of phase IV is manifested as the unequal population on the $\pm1$ components. In the phase V, where the rotational symmetry is broken, an angular striped phase is observed. Therefore, we conclude that the main physics predicted by our 2D calculation are qualitatively unchanged in a 3D BEC system.}